\documentclass{aa}
\usepackage{amssymb}
\usepackage{multirow}
\usepackage{rotating}
\usepackage{ctable}
\usepackage{morefloats}
\usepackage{graphicx,epsfig,fancyhdr,psfig,rotating,amsmath,epsf,txfonts,natbib,epstopdf,multirow}
\usepackage[colorlinks, citecolor=blue]{hyperref}
\newcommand{\Sec}[1]{Section~\ref{#1}}
\newcommand{\Tab}[1]{Table~(\ref{#1})}
\newcommand{\Fig}[1]{Fig.~\ref{#1}}

\begin{document}

\title{Magnetic topological analysis of coronal bright points}

\author{K. Galsgaard\inst{1} \and M.~S. Madjarska\inst{2} \and F. Moreno-Insertis\inst{3,4} \and Z. Huang\inst{5} \and T. Wiegelmann\inst{2} }

\offprints{kg@nbi.ku.dk}
  
\institute{
Niels Bohr Institute, Geological Museum, {\O}voldgade 5-7, 1350 Copenhagen K, Denmark
\and
Max Planck Institute for Solar System Research, Justus-von-Liebig-Weg 3, 37077, G\"ottingen, Germany
\and
Instituto de Astrofisica de Canarias, 38200 La Laguna,  Tenerife,  Spain, fmi@iac.es
\and
Dept.~of Astrophysics, Universidad de La Laguna, Tenerife, Spain
\and
Shandong Provincial Key Laboratory of Optical Astronomy and Solar-Terrestrial Environment, Institute of Space Sciences, Shandong University, Weihai, 264209 Shandong, China
}

\date{Received date, accepted date}

\abstract
{We report on the first of a series of studies on coronal bright points investigating the physical mechanism that generates these phenomena.}
{The aim of this paper is to understand the  magnetic-field structure  that hosts the bright points.}
{We use longitudinal magnetograms taken by the Solar Optical Telescope with the Narrowband Filter Imager. For a single case, magnetograms from the Helioseismic and Magnetic Imager were added to the analysis. The longitudinal magnetic field component is used to derive the potential magnetic fields of the large regions around the bright points. A magneto-static field extrapolation method is tested to verify the accuracy of the potential field modelling. The three dimensional magnetic fields are investigated for the presence of magnetic null points and their influence on the local magnetic domain.}
{In 9 out of 10 cases the bright point resides in areas where the coronal magnetic field contains an opposite polarity intrusion defining a magnetic null point above it. It is found that X-ray bright points reside, in these 9 cases, in a limited part of the projected fan dome area, either fully inside the dome or expanding over a limited area below which typically a dominant flux concentration resides. The 10th bright point is located in a bipolar loop system without an overlying null point.}
{All bright points in coronal holes and two out of tree bright points in quiet Sun regions are seen to reside in regions containing a magnetic null point.  An yet unidentified process(es) generates the BPs in specific regions of the fan-dome structure.
}

\keywords{Sun: corona - Sun: chromosphere - Sun: activities - Methods: observational}
\authorrunning{Galsgaard et al.}

\maketitle
\section{Introduction}

The term `coronal bright point' (CBP) describes a phenomenon in the solar
atmosphere that appears in extreme-ultraviolet (EUV) and X-ray images as a small-scale multi-loop system of enhanced coronal emission that is associated with magnetic bipolar features \citep[e.g.,][]{1977SoPh...53..111G, 1993SoPh..144...15W, 2001SoPh..201..305B, 2016ApJ...818....9M}. CBPs were detected for the first time in soft X-ray photographs taken during rocket missions in 1968--1973 \citep{1973SoPh...32...81V} and later were analysed in great detail during the \textit{Skylab} mission \citep[][and the references therein]{1976SoPh...49...79G, 1976SoPh...50..311G, 1992AnGeo..10...34H}. \citet{1990ApJ...352..333H} found that CBPs show no difference of their properties in coronal holes (CH) and the quiet Sun (QS), which led the authors to conclude that they do not depend on the structure of the surrounding background corona.

CBPs have sizes in the range 10\arcsec--50\arcsec. Their average lifetime derived from observations taken with the Fe~{\sc xii}~195~\AA\ filter of the Extreme-ultraviolet Imaging Telescope (EIT) on board the Solar and Heliospheric Observatory (SoHO) is 20 hours \citep{2001SoPh..198..347Z}, while recently \citet{2016ApJ...818....9M} reported a lifetime ranging from 2.7 to 58.8 hours from a study of 70 CBPs observed with  Fe~{\sc xii}~193~\AA\ filter of the Atmospheric Imaging Assembly (AIA) on board the Solar Dynamic Observatory (SDO). In X-rays CBPs have lifetimes of only 8 hrs \citep{1974ApJ...189L..93G}. \citet{2001SoPh..198..347Z} concluded that the temperatures of BPs are generally below 2$\times$10$^6$ K, which also explains their smaller size and shorter lifetime in X-rays compared to lower temperature observations. \citet{1990ApJ...352..333H} found that simultaneously measured peaks of emission in six different lines (with a large range of formation temperatures from chromospheric to coronal) were not always co-spatial, implying that the BPs may consist of a complex of small-scale loops at different temperatures and heights.  \citet{2012ApJ...757..167K} investigated the multi-thermal nature of
EUV BPs using the coronal  171, 193 and 284~\AA\, and the 304~\AA\ (chromosphere --  transition region) filter images from observations with the Extreme-UltraViolet Imager (EUVI) on board the twin STEREO satellites. The correlation coefficient between the different channels made the authors to conclude that
BPs at 171, 195, and 284~\AA\  belong to the same loop system while the BP emission in 304~\AA\ can be interpreted as coming from  cool legs of the loops.

The coronal emission evolution of CBPs strongly correlates with the variation of the total unsigned photospheric magnetic flux of the associated magnetic bipolar features (MBFs) \citep[e.g.,][]{1999ApJ...510L..73P, 2003A&A...398..775M, 2004A&A...418..313U}. \citet{2016ApJ...818....9M} concluded that the formation of the MBFs associated with CBPs involves the processes of flux emergence, convergence, and local coalescence of the magnetic fluxes, while the formation of MBFs may involve more than one of these processes. Observational magnetic cancellation, that is defined as an empirical description of the disappearance and/or decrease of magnetic features of opposite polarity while they visibly interact with each other as observed in magnetograms, was found in all 70 analysed BP cases. The MBFs were found to evolve in three typical manners, namely, ``between a MBF and small weak magnetic features, within a MBF with the two polarities moving toward each other from a large distance, and within a MBF whose two main polarities emerge in the same place simultaneously''. 

\citet{1987ARA&A..25...83Z} suggested two possible scenarios that can explain the observed magnetic cancellation. One is a simple submergence, when a pre-existing loop descends into the convection zone. Another scenario requires magnetic reconnection either above or below the photosphere, which is named reconnection submergence. Reconnection submergence is discussed in detail by \citet{1989ApJ...343..971V}. From simultaneous measurements of the magnetic field in the photosphere and chromosphere \citet{1999SoPh..190...35H} concluded that the magnetic flux ``is retracting below the surface for most, if not all, of the cancellation sites studied''. Furthermore, \citet{1994ApJ...427..459P} developed a model of BPs based on converging motions of magnetic features which can trigger magnetic reconnection and thereby energize a BP. The model was further developed in three-dimensions by \citet{1994SoPh..153..217P} and tested by two-dimensional numerical experiments by \citet{2006MNRAS.366..125V, 2006MNRAS.369...43V}. 

Reconstructing the three dimensional (3D) magnetic topology of CBPs is essential in modelling these or any other solar phenomena. \citet{1994SoPh..151...57P} were the first to derive the magnetic topology of two observed CBP.  They used a low number of point sources to capture the 3D structure of the magnetic field. This showed that the BP structure aligned with the separator line connecting two 3D null points located in the photosphere. \citet{2008A&A...492..575P} used the MPOLE topology code of \citet{1996SoPh..169...91L}, that is based on reducing a full magnetogram to a distribution of monopoles in the photospheric surface, to calculate the 3D potential magnetic field of one BP and analyse the skeleton of the magnetic field. The authors found a very close agreement between the calculated magnetic field configuration and some of the X-ray loops composing the studied bright point. This made them suggest that ``a large fraction of the magnetic field in the bright point is close to potential''. \citet{2011A&A...526A.134A} applied the same method to another BP with the same main conclusion. \citet{2012ApJ...746...19Z} extrapolated a potential field model to derive the magnetic topology of 13 CBPs using Global Oscillation Network Group (GONG) Big Bear and Learmonth magnetograms. Only two of their BPs were found to be associated with the so-called ``embedded bipolar region'' where one polarity is surrounded by another polarity. This resulted, for both cases, in a coronal magnetic null point defining a dome-shaped separatrix surface. The remaining 11 BPs were found to be associated with simple bipolar regions creating loop systems, as found in \citet{2008A&A...492..575P} and \cite{2011A&A...526A.134A}.

The aim of the present paper is to investigate the magnetic skeleton of CBPs
in both CH and QS regions using a long time-series of high resolution longitudinal magnetograms from the Solar Optical Telescope (SOT) on board the Hinode satellite combined with X-ray images taken with the X-ray Telescope (XRT) on board the same satellite. Knowing the magnetic skeleton of the BP regions is important for determining the possible mechanisms that are responsible for heating the plasma to coronal temperatures. On a longer time scale this allows us to perform time dependent 3D modellings to further enhance our understanding of the dynamical evolution of basic BP magnetic field structures. 
In \Sec{obs.sec} we describe the observations. The results are reported in
\Sec{res.sec}. Discussion and conclusions are given in \Sec{disc.sec} and \Sec{conc.sec}.

\section{Observations, modelling and analysis methodologies}
\label{obs.sec}

\subsection{Observational material}
The observations analysed in the present study were taken by XRT \citep{2007SoPh..243...63G} and SOT \citep{2008SoPh..249..167T} on board the Hinode satellite. Five datasets obtained in QS and equatorial CH regions were selected and 10 bright points were identified for further analysis from which 7 were located in the CH regions and 3 in QS regions. Details on the XRT and SOT data can be found in \Tab{tab:bp_obs}. The CH and one of the QS (on 2007 November 27) XRT observations were taken with the Al$\_$Poly filter. The QS data on 2007 October 10 were obtained with both the C$\_$poly and Al$\_$poly filters, but only the C$\_$poly filter data were used for further analysis because of their better cadence rate. The temperature response of both filters peaks around 8$\times$10$^6$~K with the Al$\_$poly filter having a stronger response at lower temperatures. The pixel size of the X-ray images is 1\arcsec $\times$ 1\arcsec. The SOT observations represent series of Stokes V and I images taken with the Narrowband Filter Imager (NFI) in the Na~{\sc i} D 5896~\AA\ spectral line which provide measurements of the magnetic field in the chromosphere \citep{2008SoPh..249..233I}. The magnetograms have a pixel size of 0.16\arcsec $\times$ 0.16\arcsec. The data calibration and alignment are performed as described in \citet{2012A&A...548A..62H}. The QS SOT data on 2011 June 18 were taken after the launch of the Solar Dynamic Observatory (SDO) and, therefore, full-disk solar observations were available in the extreme-ultraviolet (EUV) obtained with the Fe~{\sc xii}~193~\AA\ filter of the Atmospheric Imaging Assembly \citep[AIA,][]{2012SoPh..275...17L} at 12~s cadence and 0.6\arcsec\ $\times$ 0.6\arcsec\ per pixel spatial resolution. To complete the SOT dataset that only had 9 out of 19 good quality magnetograms, we also used data from the Helioseismic and Magnetic Imager (HMI) on board SDO that have 45~s cadence and 0.5\arcsec $\times$ 0.5\arcsec\ pixel size resolution. From the observed regions given in \Tab{tab:bp_obs}, we have chosen the bright points that are isolated showing a simple structure and are located sufficiently far from the domain boundaries to allow for a meaningful magnetic field extrapolation.   The typical error in the alignment between the magnetogram and XRT data is on the order of 2\arcsec\ and this error dominates over the projection effect due to the height difference in the two types of observations. This effect is most pronounced for the first dataset  (see Table~\ref{tab:bp_obs}). For the 3D visualisations shown below, the aligned low resolution X-ray data have been interpolated to the associated magnetogram resolution using the idl procedure {\it interpolate} including the keyword {\it cubic=-0.5}.

\begin{table*}[!ht]
\centering
\caption{\label{tab:bp_obs} \textbf{Data used in the present study.}}
\begin{tabular}{l c c c c c c c }
\hline
\hline
Date &Time & {Cadence (s)} & \multicolumn{2}{c} {X-cen, Y-cen} & SOT FOV & XRT FOV & BP No. \\
& (UT) & XRT/SOT & XRT & SOT & &\\
\hline
2007-Nov-09 & 06:38--14:59 & 20/30 & -540\arcsec,-222\arcsec&  -499\arcsec, -200\arcsec&261\arcsec\ $\times$ 148\arcsec & 366\arcsec $\times$ 366\arcsec & 1, 2, 3\\
2007-Nov-12 & 01:21--10:58 & 40/90 &  -47\arcsec,-228\arcsec&-7\arcsec,-206\arcsec &276\arcsec\ $\times$ 163\arcsec & 394\arcsec $\times$ 394\arcsec & 4, 5, 6, 7\\
2007-Oct-10 & 18:47--23:58 & 30/60 & -214\arcsec,-70\arcsec & -174\arcsec, -39\arcsec &  276\arcsec\ $\times$ 164\arcsec&394\arcsec $\times$ 263\arcsec& 8 \\
2007-Nov-27 & 14:00--17:57 & 30/60 & 263\arcsec,-42\arcsec &  112\arcsec,-59\arcsec& 225\arcsec\  $\times$ 113\arcsec & 394\arcsec $\times$ 263\arcsec& 9 \\
2011-June-18 & 14:16--15:13 & --/180 &--&  7\arcsec, 46\arcsec & 225\arcsec\ $\times$ 113\arcsec & --  &10 \\ 
\hline
\end{tabular}
\end{table*}

\begin{table}[!ht]
\centering
\caption{\label{bp_relation.tab} \textbf{Bright point references relative \citet{2012A&A...548A..62H}}}
\begin{tabular}{c c}
\hline
\hline
Label in Huang et al. 2012 & Label in this paper \\
\hline
Huang 1, 9 Nov.& BP1 \\
Huang 4, 9 Nov.& BP2 \\
Huang 3, 9 Nov.& BP3 \\
Huang 2, 12 Nov.& BP4 \\
Huang 1, 12 Nov.& BP5 \\
Huang 8, 12 Nov.& BP6 \\
Huang 10, 12 Nov.& BP7 \\
\hline
\end{tabular}
\end{table}

\subsection{Potential and Magneto-static modelling}
\label{extrapolation.sec}
The field of view of the SOT data given in \Tab{tab:bp_obs} was used,  at  full observational data resolution, as the photospheric boundary condition for a potential field extrapolation. This approach uses a Fast-Fourier Transform (FFT) algorithm to solve for the scalar potential 
and a 6th-order centred finite difference operator to derive the 3D magnetic field. The FFT approach requires the magnetogram data to be 2D periodic. This is achieved by smoothing the magnetogram data to zero along the domain edges using a hyperbolic tan function centred 10 pixels inside the domain, and with a halfwidth of 3 pixels. This alteration to the observational data is not critical for the extrapolated magnetic field as most of the investigated BP regions are located with a sufficient distance to the boundaries not to be significantly influenced by this change. The extrapolations are solved to a height equal to one quarter of the shortest magnetograms' sizes (except for BP9 that is much higher, see the discussion of BP9 below). 

\begin{figure}[!ht]
{
\includegraphics[scale=.5]{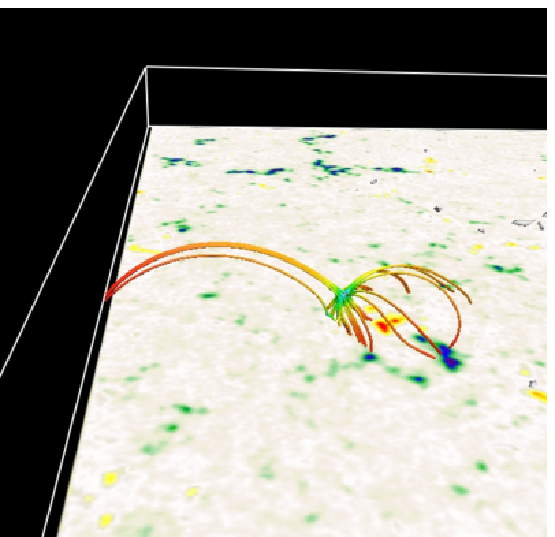}
\hfill
\includegraphics[scale=.5]{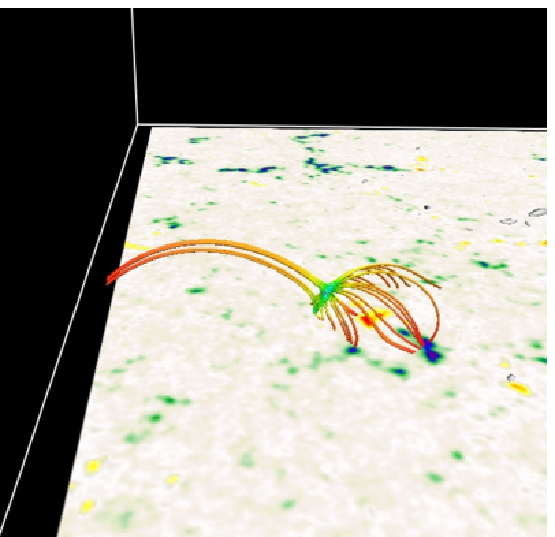}
\hfill
\includegraphics[scale=.5]{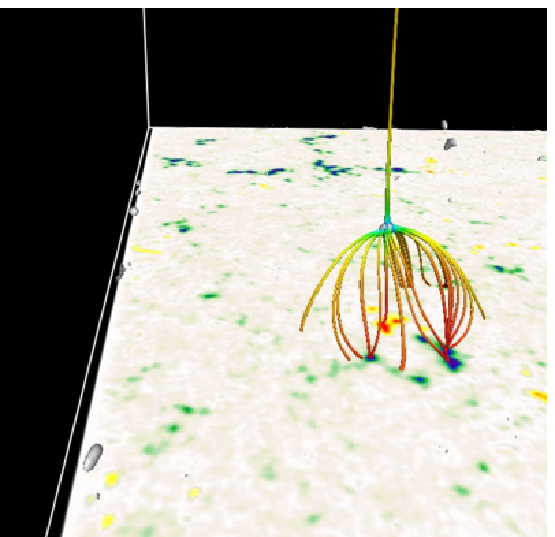}
}
\caption[]{Two extrapolations of the magnetic field using a skirt of 10 pixels around the original magnetogram to enforce flux balance and one extrapolation without the skirt. Left -- the MS model extrapolation, middle -- the potential extrapolation and right -- potential extrapolation of the original data  with the boundary smoothing to remove the signal to enforce a periodic setup. The field lines represent the magnetic field line structure around the null point associated with the BP1 case (see the discussion below).
\label{extrapolation_compare.fig}}
\end{figure}

A magneto-static (MS) field extrapolation \citep{2015ApJ...815...10W} was also used to test the reliability of the potential solution. The MS approach is based on a special class of linear MS equilibria as proposed by \citet{low91}. The electric current density is described as
\begin{equation}
\nabla \times {\bf B } = \alpha_0 {\bf B } + f(z) \nabla B_z \times {\bf e_z},
\label{lin_mhs}
\end{equation}
where $\alpha_0$ is the linear force-free parameter controlling field aligned currents and $f(z)$ is a function which controls horizontal currents. {In the following $f(z)=a exp(-k z]$ is used and the two constants have the following values; $\alpha_0=1$ and $a=1$.}  The equilibria are solved with the help of a FFT method. 
The MS modelling requires the input data to be in magnetic flux balance. The first dataset of the CH regions obtained on 2007 November 9 was used for the comparative test (see \Tab{tab:bp_obs}). The region is dominated by open magnetic field, and therefore, has a clear flux imbalance. To fulfil the flux balance requirement of the MS approach, a band of 10 pixels of uniform flux is placed around the observational data. These adjustments of the new magnetogram make clear alterations to the derived field line connectivity inside the domain, while still allowing the locally imbalanced flux to reach into the atmosphere before connecting down to the fictive flux region ``around'' the original data. 

Both a MS and a potential extrapolation were derived for the modified magnetogram. The field line structure of the BP1 region found in the MS (left frame) and the potential (middle frame) extrapolations are shown in \Fig{extrapolation_compare.fig}. This demonstrates that the two extrapolation methods provide magnetic field structures that, for this region, are comparable. {This indicates, as also stated by \cite{}, that the magnetic field is close to being potential for the BP regions. For a closer comparison of the two solutions the characteristics of the central null point in the figure has been investigated.}  The position and eigenvalues of the null point for the different extrapolation approaches are given in \Tab{null_compare.tab}. This shows that the two extrapolation methods produce comparable magnetic field structure. For a further comparison, a field line trace (right frame) in \Fig{extrapolation_compare.fig}, and the null point information is also provided for a potential extrapolation of the original data with  the smoothing described above to provide a periodic dataset.
A significant deviation is only found for the height of the null point that is clearly higher in the original data. The differences are caused by the deviations from flux balance, which for the original data allow the magnetic field to expand higher into the atmosphere.  In the following investigation, only the potential field extrapolation on the original datasets is used.

\begin{table*}[!ht]
\centering
\caption{\label{null_compare.tab} Parameters of the BP1 associated null point using two different extrapolations methods with imposed and without flux balance.  The horizontal positions are given in gridpoints relative to the lower left corner of the SOT data set, where the pixel size is equivalent to the SOT magnetogram resolution of 0.16\arcsec. The height is measured in the same units.}
\begin{tabular}{c c c}
\hline
\hline
Extrapolation method and field & Position (grid points) & Eigenvalues \\
\hline
Mag-static-skirt   & (1004.7, 839.3, 16.4) & (-0.39, 0.18, 0.21) \\
Potential-skirt    & (1004.7, 839.4 , 15.8) & (-0.43, 0.23, 0.20) \\
Potential          & (1002.2, 812.9, 36.5) & (-0.33, 0.15, 0.18) \\
\hline
\end{tabular}
\end{table*}

\subsection{Magnetic topology analysis methods}\label{top.sec}

The extrapolated 3D potential magnetic fields were investigated for magnetic null points using the method of \citet{2007PhPl...14h2107H}. This approach employs a trilinear method to identify null points in a 3D gritted magnetic field dataset. The method works successfully when the typical length scales of the magnetic field are clearly larger than the grid resolution. As the high frequency information in the FFT extrapolation decays exponentially with wave number and height, this condition is typically fulfilled above a given height in the atmosphere (5-10 horizontal grid points). This may further be tested by calculating the $\nabla \cdot B$ through summing up the eigenvalues of the null point. If this requirement is satisfied numerically, then the linear criterion is typically sufficiently fulfilled.

To identify the magnetic field structures at the null locations, their characteristics are derived solving for the eigenvalues of the associated Jacobian matrix of the magnetic field \citep{1997GApFD..84..245P}.  The nullfinder approach provides a very high number of identified 3D null points inside the domain. Most of these are located very low in the atmosphere, as also found in \citet{2009SoPh..254...51L}, and a fair number of these are associated with the imposed boundary smoothing of the magnetic field.  Only null points above a given minimum height (10 gridpoints) have been considered. To qualify as the null associated with a given BP region, they have to be located sufficiently close to the BP position. Typically, only one null point fits this description and its time evolution is easy to follow and is investigated below.

The 3D magnetic field is visualised using the VAPOR software to further investigate the magnetic field topology around the locations of the identified BPs. This analysis reveals that 9 of the 10 BPs in the sample contain a dominating magnetic null point located in the coronal domain. The null points have only real eigenvalues \citep{1997GApFD..84..245P}. The general structure of all BP null points is such that the spine axis (associated with the dominating eigenvalue and its eigenvector), in one direction, connects to the minority polarity below the null point and, in the opposite direction, connects towards the top boundary of the domain. In comparison, the fan planes are asymmetric with the main axis (dominating fan eigenvalue and vector) pointing towards the nearby dominating flux concentration. The presence of the null above the minority polarity flux concentration implies that all flux from the minority polarity closes back to the photosphere, while the ambient excess flux  connects to the upper boundary.

\section{Results: data analysis and modelling}
\label{res.sec}

The seven selected CH bright points (BP1--BP7) were previously discussed in detail in \citet[][hereafter HBPs]{2012A&A...548A..62H} (see also \url{http://star.arm.ac.uk/highlights/2012/603/bpmag}). \Tab{bp_relation.tab} relates the bright point numbering in \citet{2012A&A...548A..62H} with the numbering used in the rest of this paper. The temporal evolution of the BPs' X-ray emission and their associated magnetic polarities including magnetic flux decrease rate of BP1, BP3 and BP4 are given by \citet{2012A&A...548A..62H}. The BPs were chosen as they have well isolated magnetic field structures only related to the bright point phenomena. Also, all examples are sufficiently long lived to allow for a qualitative investigation of the magnetic field evolution during their lifetime. 

\begin{table*}[!ht]
\centering
\caption{\label{tab:bp_an}  General bright point properties.}
\begin{tabular}{l l l l}
\hline
\hline
BP No. & MBF evolution & Null point & Projected X-ray/EUV location \\
\hline
1 & Convergence \& decrease & Decreasing height & Above one of the main polarity\\
  &                                         &                            &  flux concentrations\\
2 & Convergence \& divergence. Diffuse & Moves up and down & Spread around mainly inside \\
  & flux emergence                                &                                & the dome, concentrated at the \\
  &                                                       &                                & strongest minority flux region \\
3 & Convergence & Low lying null point & Inside the fan-dome,  \\
  &                     &                               & appears during a jet\\
4 & Flux emergence and convergence of  & Rise in the corona & Inside the fan-dome, moves \\
  & emerged positive flux with pre-existing &                                     & around\\
  & negative                                            &                                      & \\
5 & Convergence of opposite flux & Decreases in height & Inside the fan-dome\\
6 & Flux convergence &  Decreases in height & Located above a dominant flux  \\  
  &                            &                      & concentration that includes a \\
  &                            &                      & fraction of the fan-dome\\
7 & Granular size motions & Moves up and down  & Inside the fan-dome \\
8 & Convergence of negative flux with & Increases in height and & In the area between the \\
   & one of the strong positive fluxes    & expands in volume        & approaching flux patches \\
9 & Convective motion and possible & Located far above the & A low lying bipolar loop system \\
   & small scale emergence              &  BP region                  & located inside the dome\\
10 & Increased flux concentrations with time        & No null identified & A bipolar loop system \\
    & and a slow convergence of opposite polarities & & \\
\hline
\end{tabular}
\end{table*}

All BPs reveal a dynamical evolution observed in X-rays/EUV (see for example Figures~6, 9 and 13 in HBPs) on time scales as short as the data cadence (see Table~\ref{tab:bp_obs}). The analysed BPs were observed during different stages of their lifetime but the long duration of the observations provides a sufficient coverage for the requirements of the present study. It is important to mention that the choice of data was entirely dictated by the quality of the magnetic field data. Presently, SOT provides the highest quality of uninterrupted long time series of magnetic field data that are crucial for modelling of small scale solar phenomena. Therefore, these data fulfil the requirements of our investigation on the magnetic field structure of BPs and the changes of their footpoint properties related to their dynamical coronal evolution.  
 This investigation focuses on the normal component magnetic data from NFI/SOT and HMI/SDO (one case) to provide the boundary condition for the potential field extrapolation. There are no available vector magnetogram that could have been used for these investigations.

This section is focused on the general changes in the photospheric magnetic field of the BPs, the specific characteristics of the magnetic field topology and the associated X-ray/EUV emission that identifies a feature as a coronal bright point. Here it has to be stressed that coronal bright points are only seen in X-rays 
for a short period of time during their entire lifetime.
This implies that they are active to this temperature level only when their magnetic structure is sufficiently stressed. BPs have a longer lifetime when observed in the lower temperature passband imaging data \citep{2001SoPh..198..347Z}, i.e. EUV. It may, therefore, be possible that some of the general dynamical changes seen in the magnetic observations do provide energy releases, but that these are not sufficient to reach high coronal temperatures. This makes it difficult to see which specific changes in the photospheric footpoint positions of the magnetic field could result in important large energy releases. For this we need much more detailed dynamical investigations using at least the same quality of magnetic field data (SOT) together with high resolution and temperature coverage imaging data, e.g. AIA/SDO and IRIS, to obtain a better understanding on this relationship. An ongoing dedicated observing campaign should provide suitable data for further analysis.

\begin{figure*}[!ht]
{
\includegraphics[scale=.8]{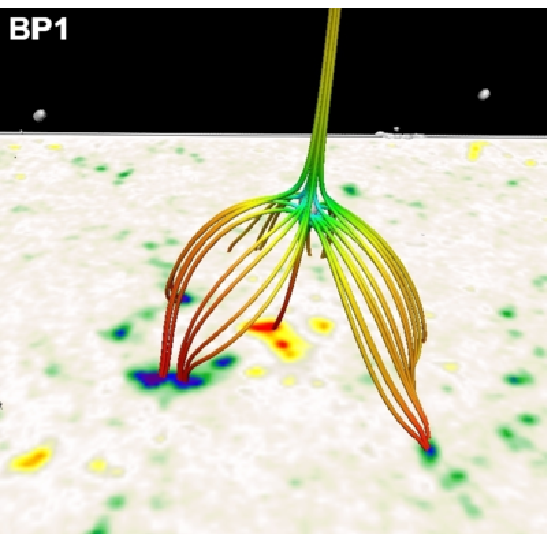}
\hfill
\includegraphics[scale=.8]{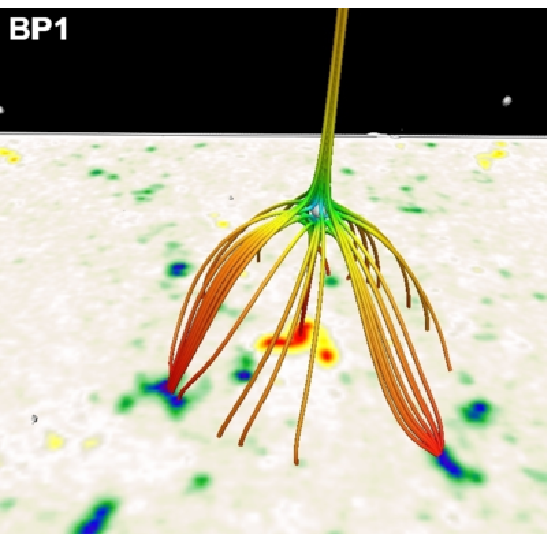}
\hfill
\includegraphics[scale=.8]{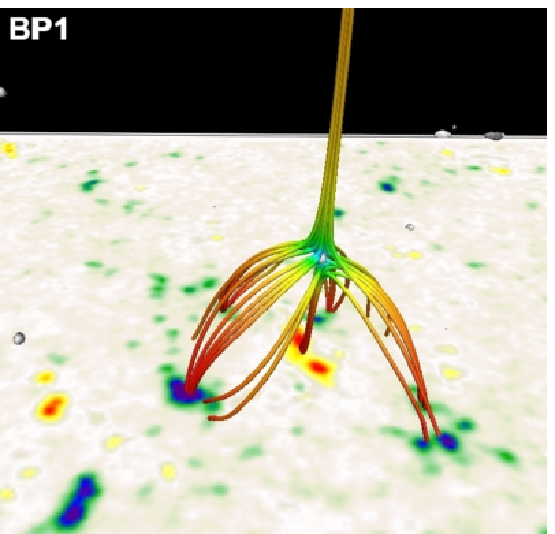}
\hfill
\includegraphics[scale=.8]{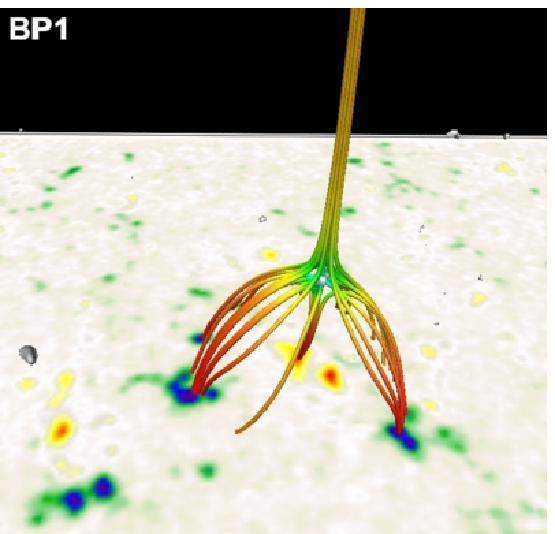}
}

\vspace{.1cm}
{
\includegraphics[scale=.8]{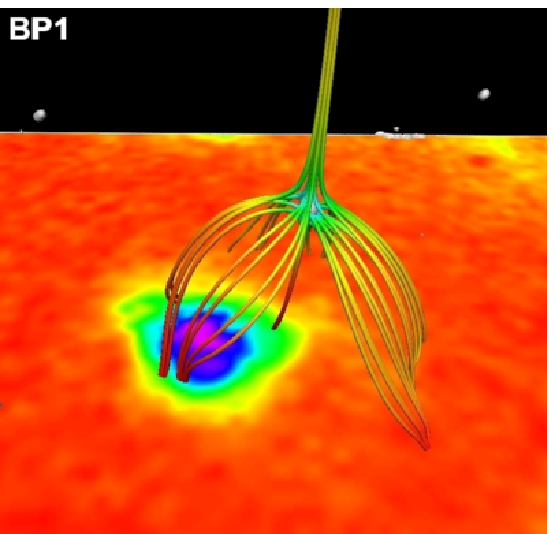}
\hfill
\includegraphics[scale=.8]{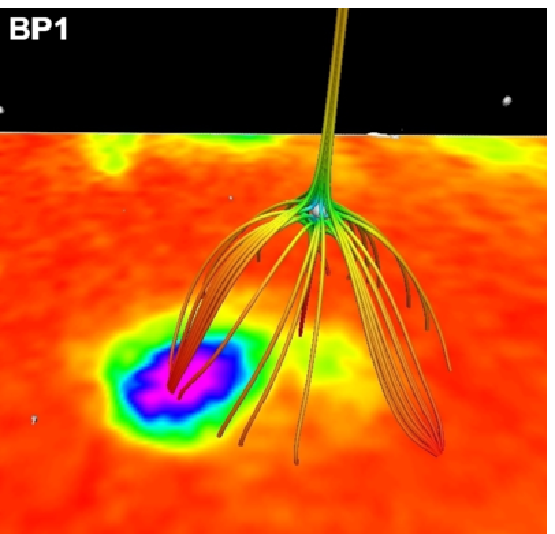}
\hfill
\includegraphics[scale=.8]{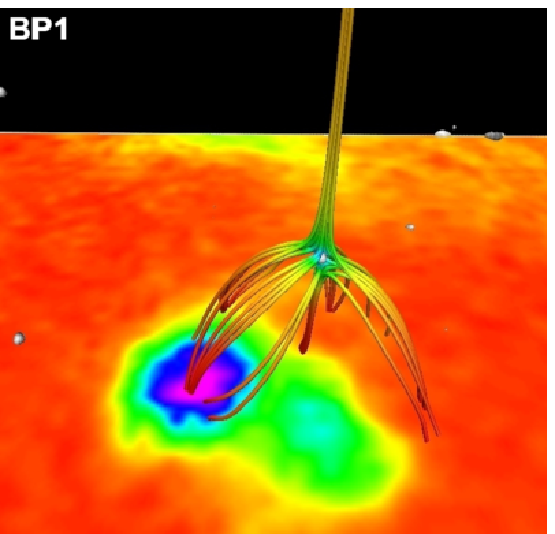}
\hfill
\includegraphics[scale=.8]{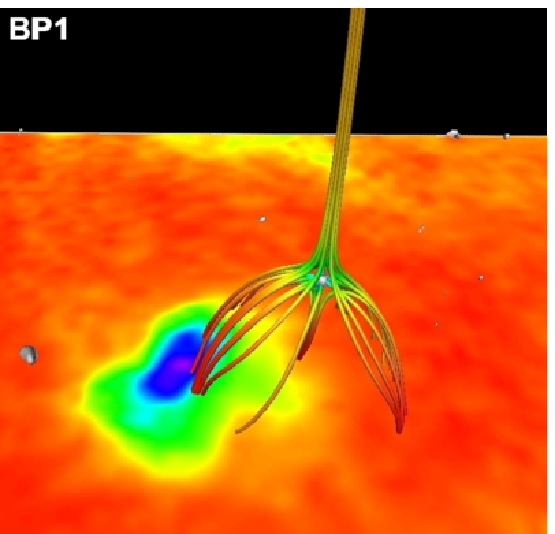}
}

\caption{Time evolution of the BP1 case. Top row shows the evolution of the magnetogram and the corresponding potential magnetic field extrapolation. The magnetogram is scaled to $\pm$200~G, where blue represent negative and red positive polarity. The grey isosurface indicates the location of the magnetic null point, with the field lines being traced from the vicinity of the null region. The field line colour represents the magnetic field strength, with red being strong, and green-blue represent the weak field. The lower row, shows the same magnetic fields, while the bottom boundary represents the XRT observations shown on a linear scale, with the red being low values and purple representing the peak values. The images correspond to the times 06:38:13~UT, 08:38:05~UT, 10:38:08~UT and 11:59:35~UT (left to right) on 2007 November 9. See the included animation for the time evolution.
\label{BP1.fig}}
\end{figure*}

{\it BP1}  (\Fig{BP1.fig}) is present from the start of the observations (at 06:38~UT) and it disappears at $\sim$12:00~UT. It was found in an equatorial coronal hole observed on 2007 November 9 with dominant negative polarity flux (hereafter dominant) and parasite or minority positive flux (hereafter minority). The XRT images show a relatively round shaped BP with a diameter of $\sim$10\arcsec\ that is preserved for most of the observing period. Impulsive variations of the X-ray emission are observed for the whole duration of the BP (see Figure~8 in HBPs), two of which at around 09:44~UT and 10:10~UT exceed more than twice the BP average brightness and last for several minutes. The BPs projected X-ray emission is located above a strong flux concentration. Small dominating magnetic fragments are constantly moving towards the minority polarity and slowly erode it away while a remote dominating polarity patch strengthens and slowly approaches the minority flux. When the BP vanishes in X-rays, the magnetic structure changes now avoiding the clear connection to the strong dominant flux concentration that previously ``hosted'' the X-ray emission (see the attached animation for the time evolution). 
\begin{figure*}[!ht]
  \center
  \includegraphics[scale=.45]{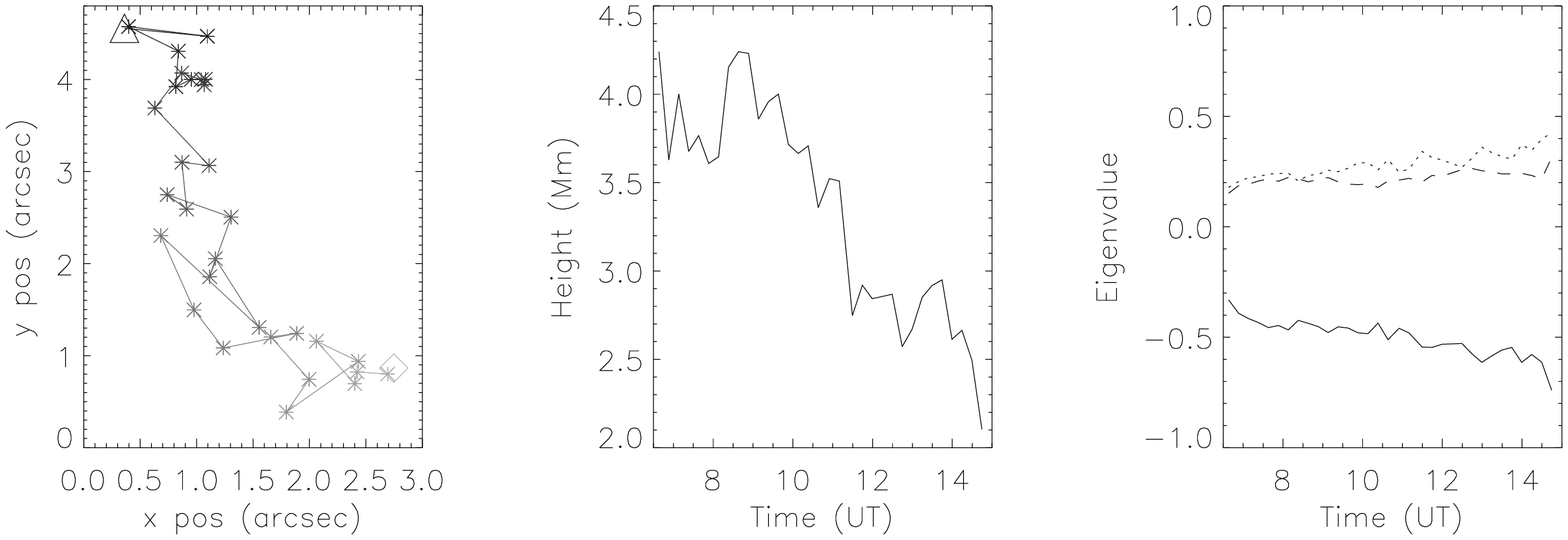}\\
  \includegraphics[scale=.45]{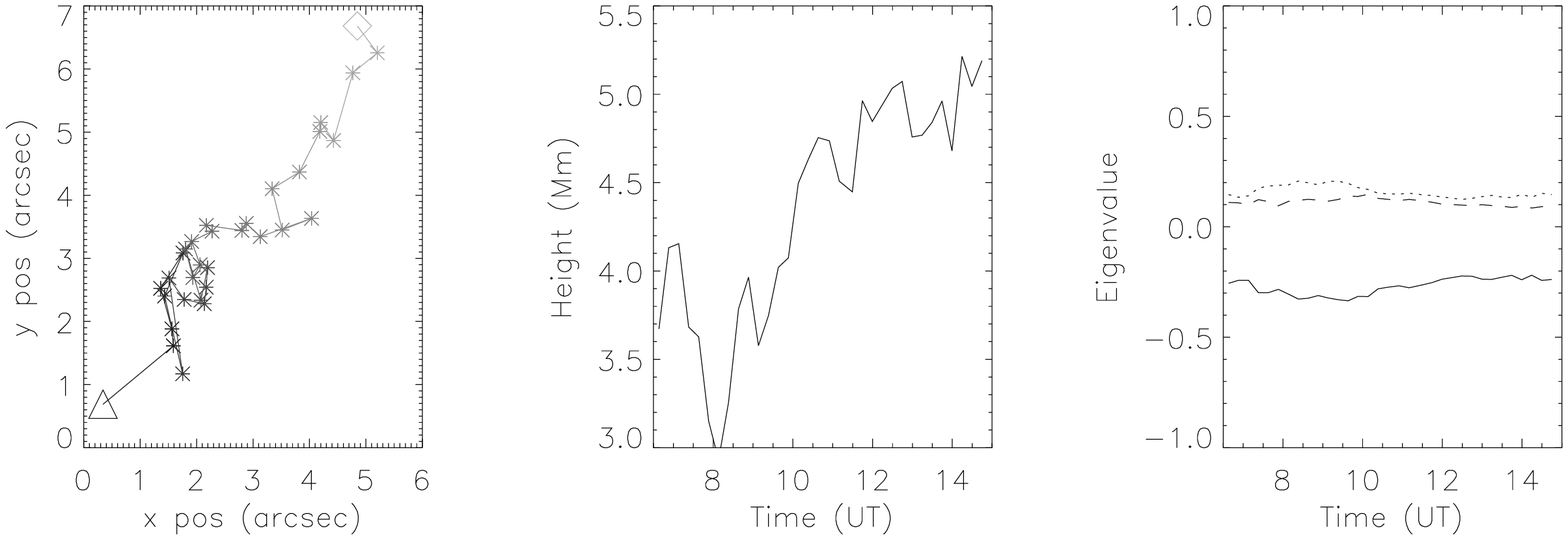}\\
  \includegraphics[scale=.45]{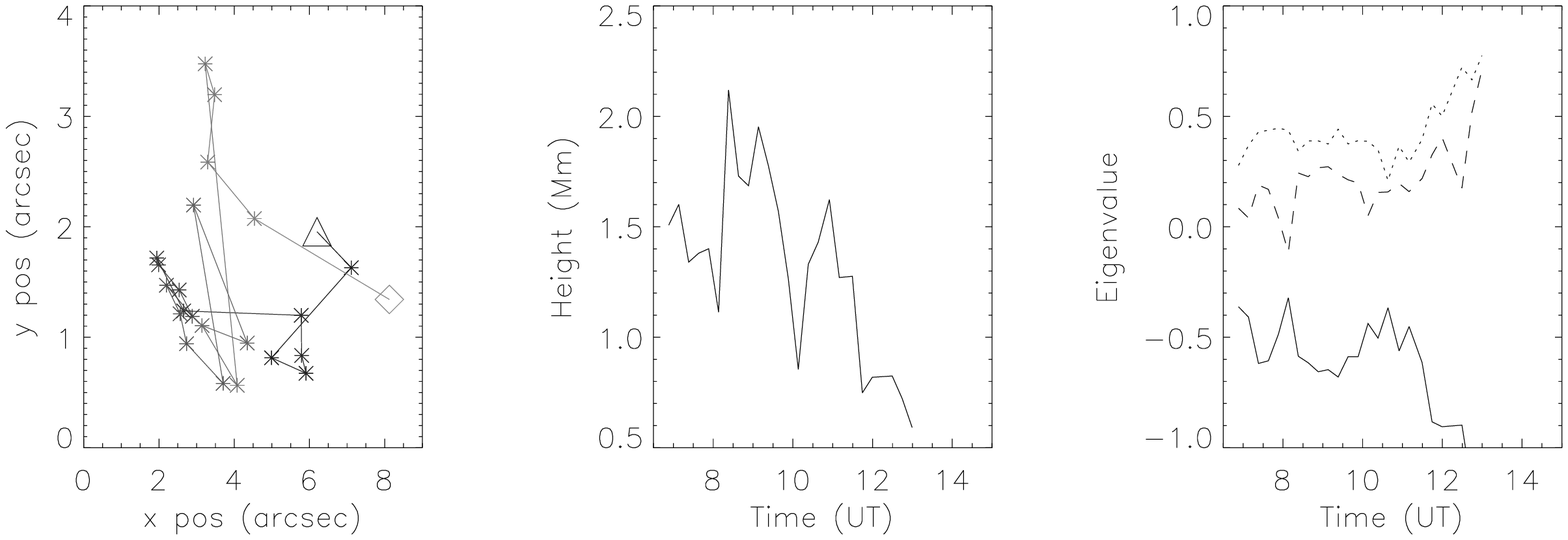}
  \caption{{\bf Left column:} Change in the horizontal null point position relative to a fictive null position for BP1 (top row), BP2 (middle) and BP3 (bottom). The grey shaded stars mark the evolution in time going from black to grey.  The first point is marked with a large triangle and the last point with a large diamond. {\bf Middle column:} Height change of the null point. {\bf Right column:} Time dependent change of the null point's eigenvalues.}
  \label{BP_1_Nulls.fig}
\end{figure*}

The temporal evolution of the BP1 null point is given in \Fig{BP_1_Nulls.fig} (top row), where the left panel shows the change in the null point horizontal position as a function of time. It reveals a jittering of the horizontal position that is mostly imposed by the spatial changes in the photospheric flux distribution and only weakly by the instrument pointing. The middle panels present the null point height as a function of time. This shows that the null point has a near monotonic decrease in height as a function of time. This motion is coupled to the general convergence of the dominating flux patches over time. It leads to a constant decrease in the total volume under the fan dome, eventually reaching a point where most of the volume is too low in the atmosphere to allow further heating of the plasma inside the dome to X-ray temperatures. Finally, the right panel shows how the three eigenvalues change in time, where it is seen that their magnitudes increase only slightly with time. This may be expected as the null point decreases in height, while the field strength and magnetic gradients in the null region increase. 

\begin{figure*}[!ht]
{
\includegraphics[scale=.8]{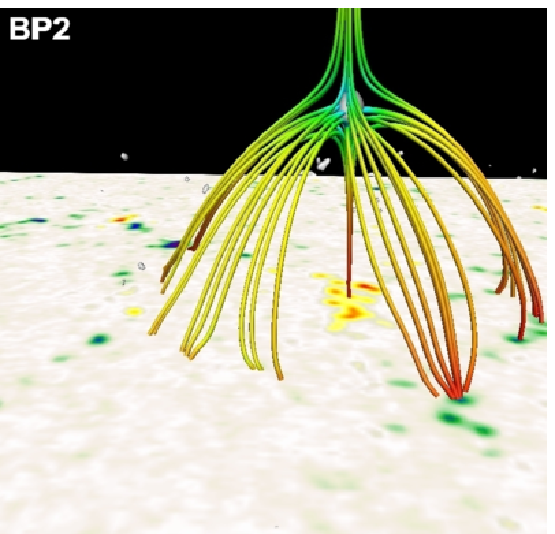}
\hfill
\includegraphics[scale=.8]{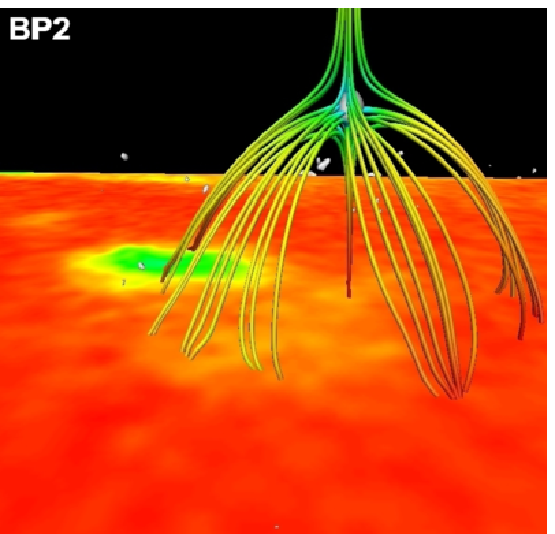}
\hfill
\includegraphics[scale=.8]{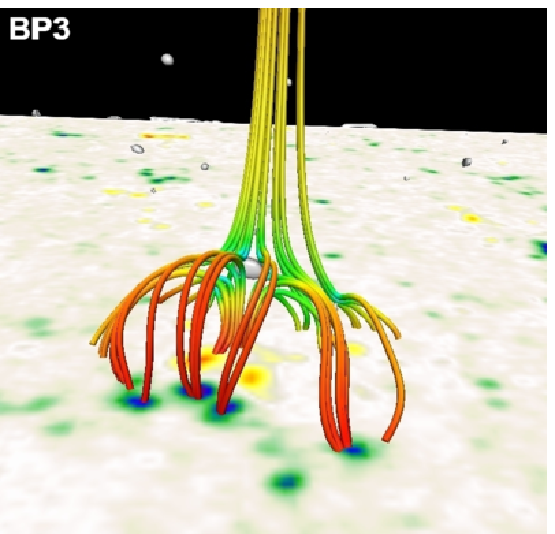}
\hfill
\includegraphics[scale=.8]{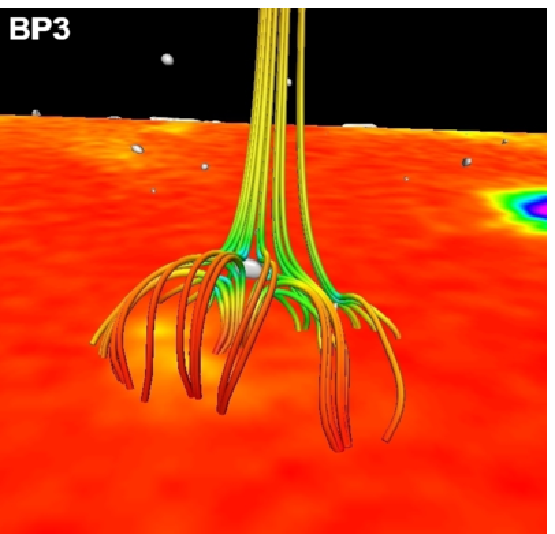}
}

\vspace{.1cm}
{
\includegraphics[scale=.8]{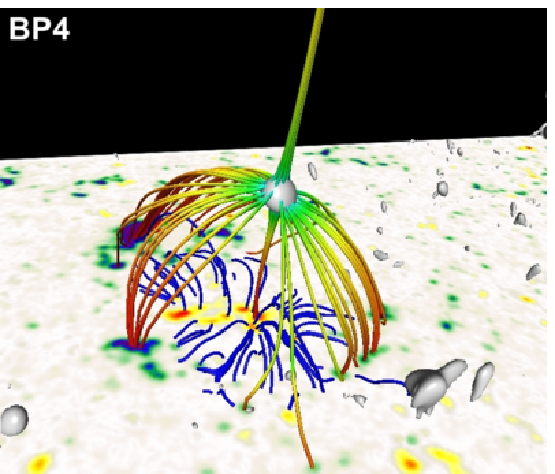}
\hfill
\includegraphics[scale=.8]{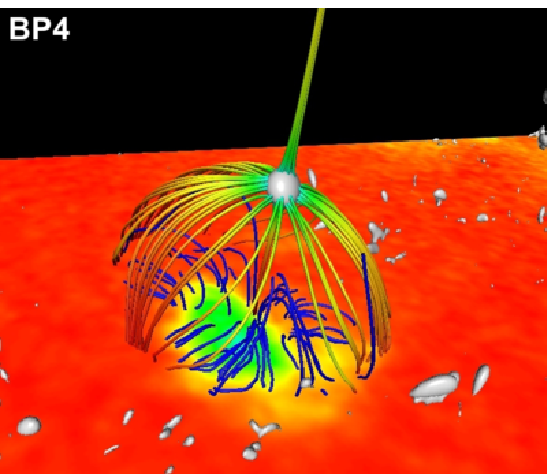}
\hfill
\includegraphics[scale=.75]{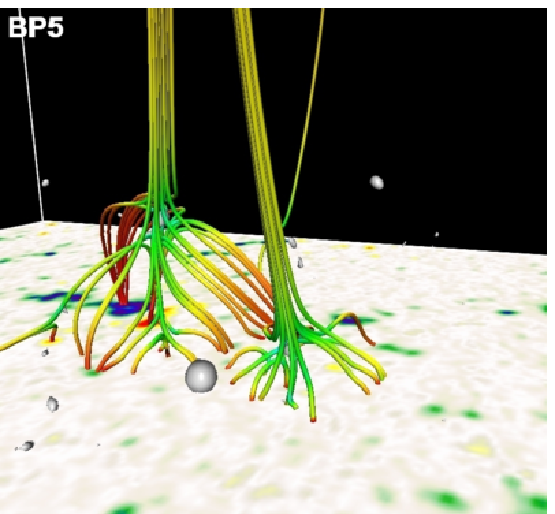}
\hfill
\includegraphics[scale=.75]{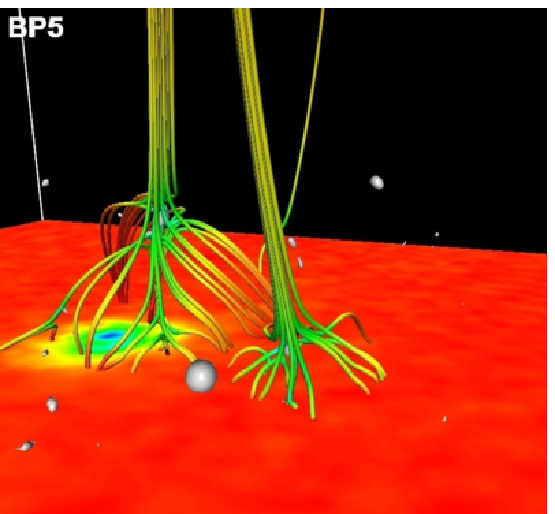}
}

\vspace{.1cm}
{
\includegraphics[scale=.75]{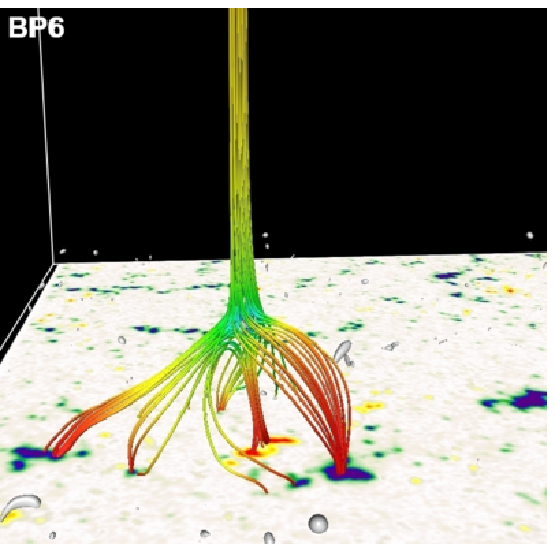}
\hfill
\includegraphics[scale=.75]{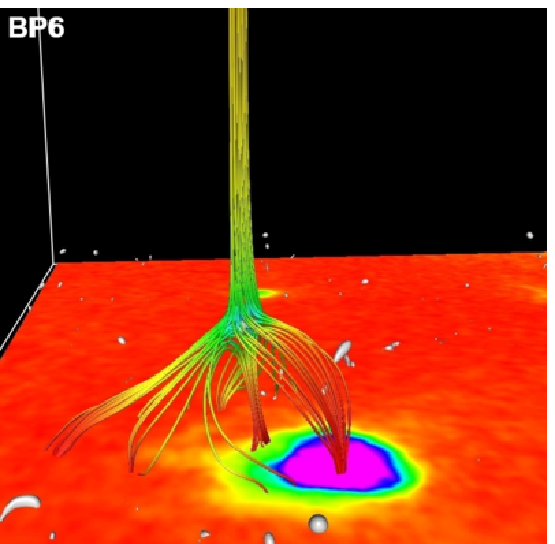}
\hfill
\includegraphics[scale=.8]{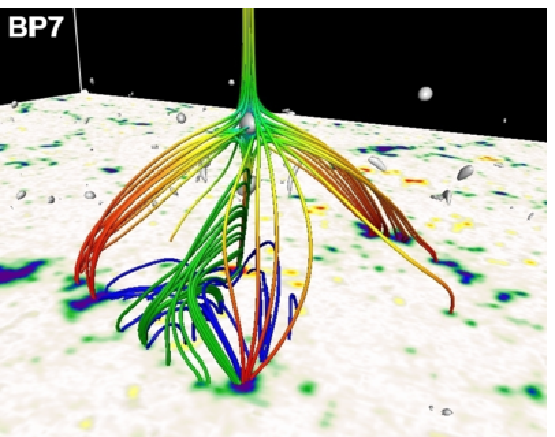}
\hfill
\includegraphics[scale=.8]{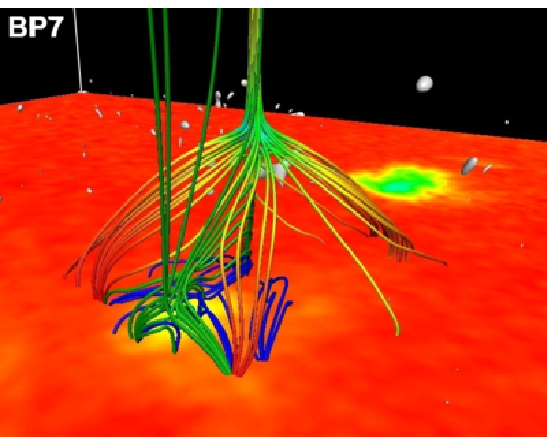}
}

\caption{Snapshots of CH bright points BP2--BP7 in pairs of two showing first the surface magnetic field followed by the XRT observations. In both cases the field lines represent the magnetic structure around the null region as in \Fig{BP1.fig}. The times for the frames are BP2  -- 14:14:36~UT, BP3 -- 10:23:07~UT, BP4 -- 06:51:03~UT, BP5 -- 06:06:03~UT, BP6 -- 05:21:35~UT and BP7 -- 08:06:05~UT on their respective dates, \Tab{tab:bp_obs}. In BP4 and BP7 additional single coloured field lines are included to show the local field line structure near the the projected  X-ray BP regions. See the included corresponding animations for their time evolution.
\label{BPrest_CH.fig}}
\end{figure*}

{\it BP2} (\Fig{BPrest_CH.fig}) becomes visible in the same CH as BP1 around 07:38~UT and it lasts for the remainder of the observations. Magnetically, this case is very different from BP1: only the minority flux is concentrated, while for most of the time the dominating polarity consists of a weak background field. The minority flux concentration increases with time following the emergence and convergence of small scale (pixel size) magnetic flux elements. During this time the dominant flux has only minor flux concentrations scattered around the minority flux. 
The X-ray intensity is lower than BP1's and it has a much more diffuse structure, with the emission ``moving'' around over a larger elongated area (around 20\arcsec $\times$ 30\arcsec), that is mainly distributed over one half of the projected dome region. It is unclear why the X-ray emission is located in  this part of the dome from just looking at the time evolution of the underlying magnetogram. Around 14:14~UT the XRT data show the highest intensity in the BP. This is associated with a period where a strong dominant polarity becomes part of the fan dome. The middle row of \Fig{BP_1_Nulls.fig} shows a near systematic advection of the null point's horizontal position in time, while at the same time the height of the null point rises in the atmosphere. This is consistent with the continued accumulation of minority flux.

{\it BP3} is also formed in the CH observed on 2007 November 9 (\Fig{BPrest_CH.fig}), and it appears in X-rays around 11:10~UT. At $\sim$11:57~UT, the BP produces an X-ray jet that lasts for around 10~min (see HBPs for more details). The BP vanishes at around 12:29~UT. The magnetic structure is dominated by a concentrated minority polarity flux pattern and a ``wall'' like distribution of dominating polarity flux on ``one'' side of the minority flux. These flux concentrations advect towards each other, which makes the magnetic structure that host the BP shrink in size.  The BP  vanishes as the minority polarity flux disappears, leading to a unipolar open field region. In the bottom row panels of \Fig{BP_1_Nulls.fig} it can be seen how the null point is moving around in the horizontal direction and its height decreases as a function of time. At the beginning of the observations the X-ray emission appears more spread out under the entire dome area, which may be explained by the divergence of the flux concentrations. As some of the flux patches form stronger concentrations, the X-ray emission concentrates above the approaching opposite polarities, i.e. in only a minor part of the dome. Before the BP appearance in X-rays, the potential field already contains a null point that is located much higher than when the BP initiates. As the BP vanishes the dome has diminished in size and shortly after it disappears.

\begin{figure*}[!ht]
  \center
  \includegraphics[scale=.45]{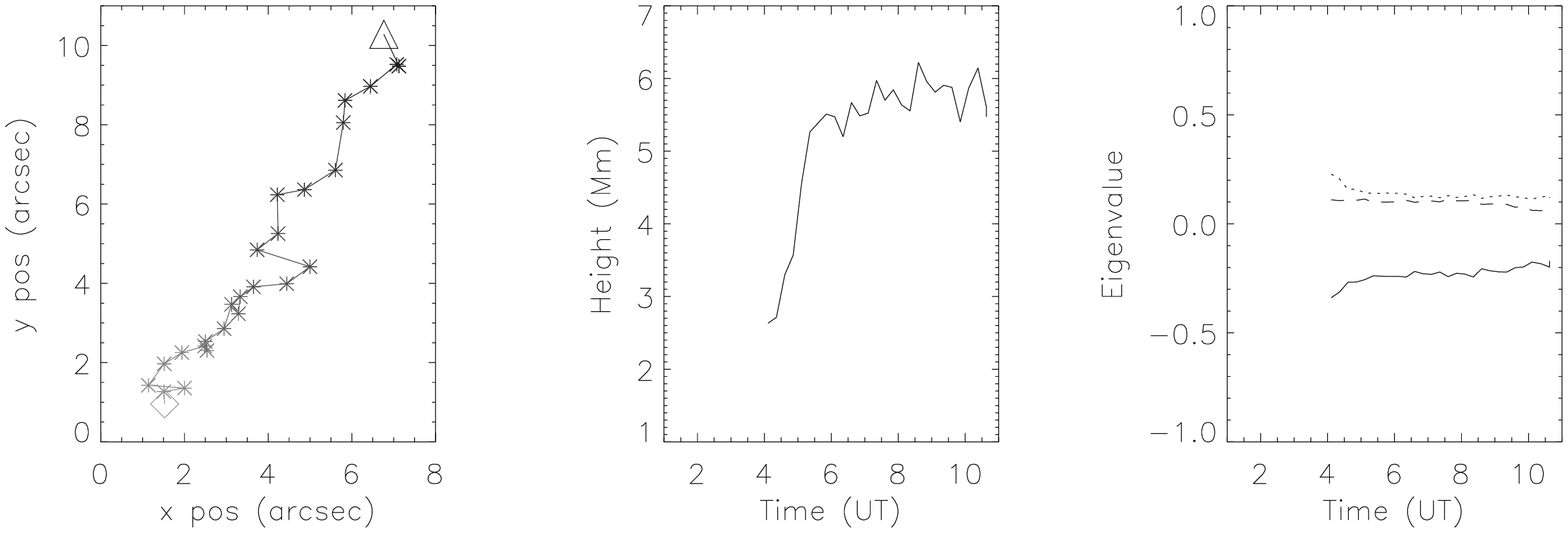}\\
  \includegraphics[scale=.45]{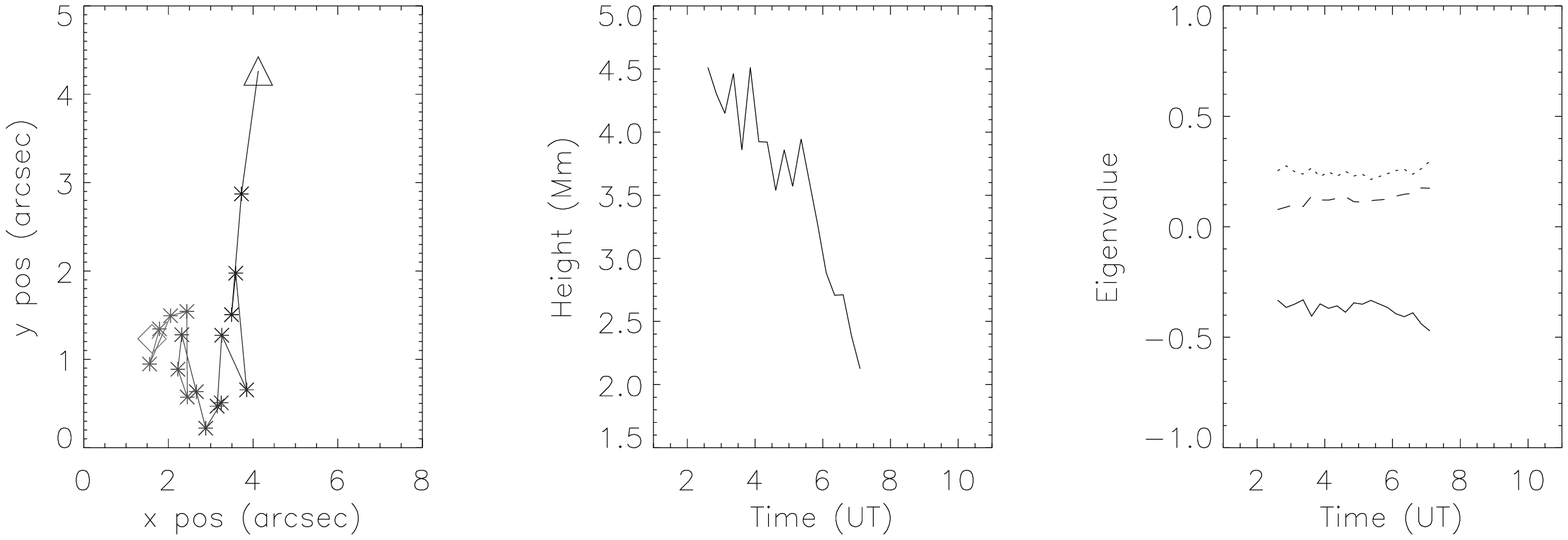}\\
  \includegraphics[scale=.45]{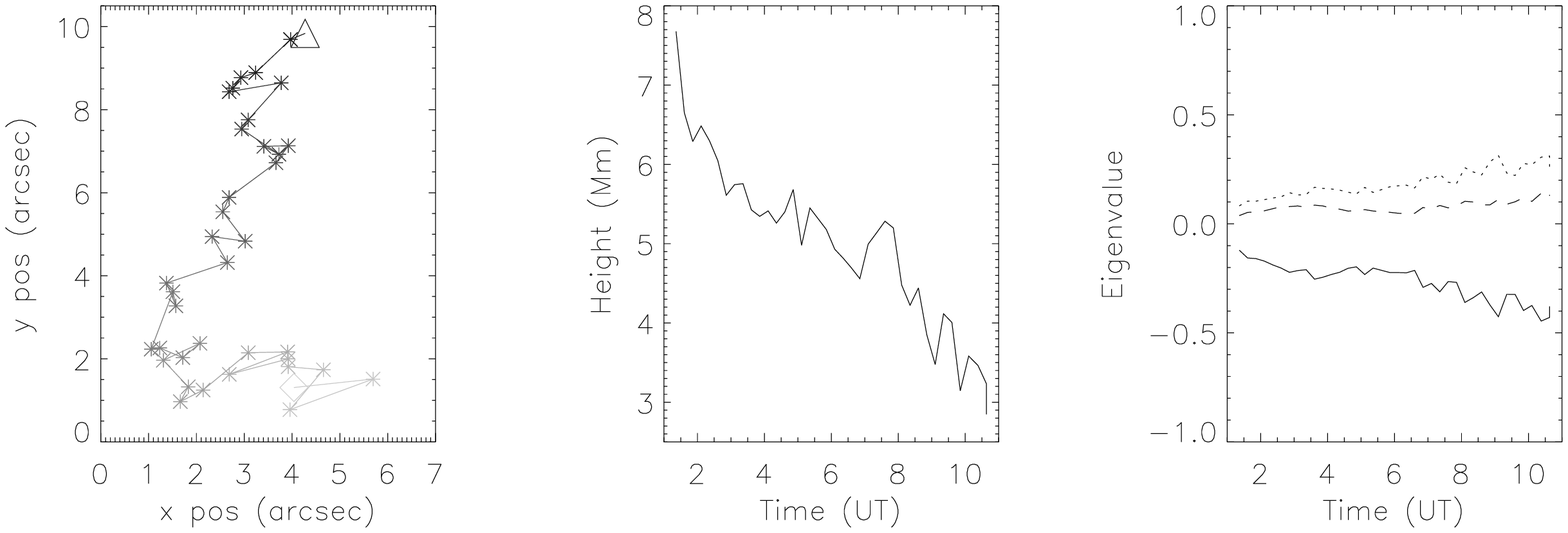}
  \caption[]{The same as \Fig{BP_1_Nulls.fig} for  BP4 (top row), BP5 (middle), and BP6 (bottom).}
  \label{BP_2_Nulls.fig}
\end{figure*}

{\it BP4} has been observed in the same CH but 3 days later on November 12 (\Fig{BPrest_CH.fig}). It becomes visible in the X-rays around 04:40~UT. Its shape changes quite dynamically but its size remains around 20\arcsec $\times$ 20\arcsec. Before the BP appears, the magnetic structure of the region is dominated by the dominant polarity, with an almost open field structure. In the period just before the BP appearance at $\sim$04:00~UT, a null point forms and starts rising in the corona (see \Fig{BP_2_Nulls.fig} (top row)). This is driven by the emergence of magnetic flux.  As the amount of minority flux increases the height of the null increases. The X-ray emission has a complex behaviour spreading over a large area which is clearly related to the complex magnetic field structure and its evolution. During the later stronger X-ray phase, the BP is located above a weak field region of opposite polarity to the stronger dominant flux concentrations. The dominant flux concentrations are enhanced during the emergence, relative to the spine footpoint. At the time of the most active X-ray phase, the amount of minority flux increases and moves in the direction of the strong X-ray emission, before the X-ray signature is finally located around the spine footpoint.

{\it BP5} is also found in the CH on November 12 (\Fig{BPrest_CH.fig}). It is present almost from the start of the observations (after 02:00~UT) and lasts until around 07:10~UT which coincides with the disappearance of the null point (\Fig{BP_2_Nulls.fig}, middle row). The BP X-ray signal shows quite some modulation with time. It has a round shape with a diameter smaller than 10\arcsec. From the start the magnetic null point is located high in the atmosphere. The height decreases with time as the minority polarity slowly approaches the dominant flux concentration and a new dominant patch is found to appear on its ``back side'' (with respect to the observer viewing direction in \Fig{BPrest_CH.fig}). Towards the end of the X-ray emission, the magnetic field complexity increases significantly, which is seen in the form of several magnetic null-point regions appearing in close proximity of the original null point at very low heights.

{\it BP6} is again a November 12 CH BP (\Fig{BPrest_CH.fig}). The BP is present throughout the whole observing period. It has almost a round shape (around 20\arcsec\ diameter) that changes to elliptic as two of the minority and one of the dominant polarities converge in time. Its intensity shows variations in time with a few major increases. The strong X-ray signal is found above the dominant flux region that occupies a small partition of the fan dome. The size of the region embedded by the fan dome is large, covering several granular cells. This is clearly seen in the time animation of the normal component of the magnetic field, that shows a jittering motion at different inter-granular lanes. Towards the end of the observations the BP decreases in size as the minority flux converges towards the largest dominating flux patch that hosted the X-ray signal. This motion leads to a decrease of the null points height, from $\sim$8~Mm to finally reaching a height of $\sim$3~Mm. This change in the magnetic character is nicely reflected in the properties of the null point, \Fig{BP_2_Nulls.fig} (bottom row). This shows both a systematic horizontal motion and the significant change 

{\it BP7} is the last case belonging to the CH observed on November 12 (\Fig{BPrest_CH.fig}). It has a magnetic null point located high in the atmosphere for the whole time of the observations (\Fig{BP_3_Nulls.fig}, top row). The dominant flux concentrations seem to be rooted in different supergranular cell corners and are with time advected around. This results in a systematic advection of the null position with, first, a systematic increase in height, followed by an equivalent decrease in height. The X-ray emission is weak for most of the time with clear fluctuations in time and a significant emission increase around 07:50~UT that lasts for approximately 20 min. The BP shape changes from almost round (10\arcsec\ across) to slightly elongated with a size of less than 10\arcsec. The maximum X-ray emission is associated with a location of generally weak magnetic flux, which for a short period of time contains a lot of small scale mixed polarity flux. During this time there is a general advection of the magnetic spine region towards the X-ray region.
\begin{figure*}[!ht]
{
\includegraphics[scale=.8]{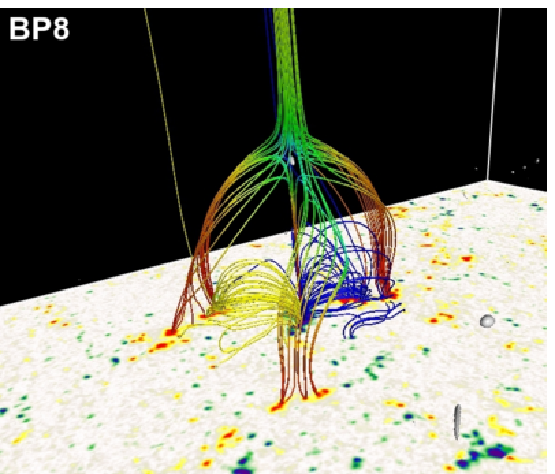}
\hfill
\includegraphics[scale=.8]{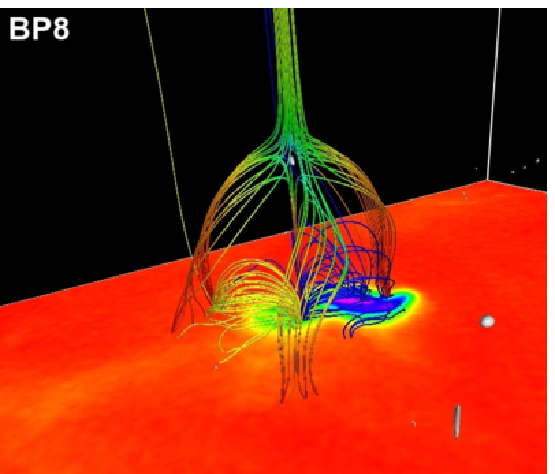}
\hfill 
\includegraphics[scale=.8]{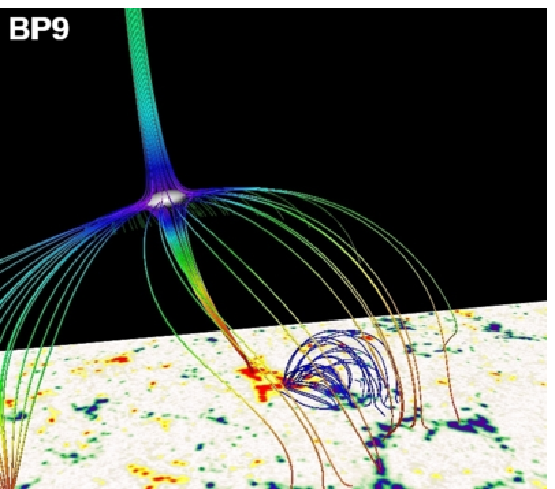}
\hfill
\includegraphics[scale=.8]{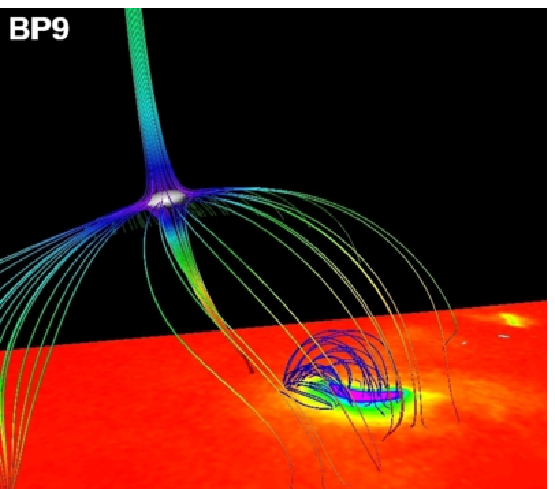}

\vspace{.1cm}
{
\includegraphics[scale=.8]{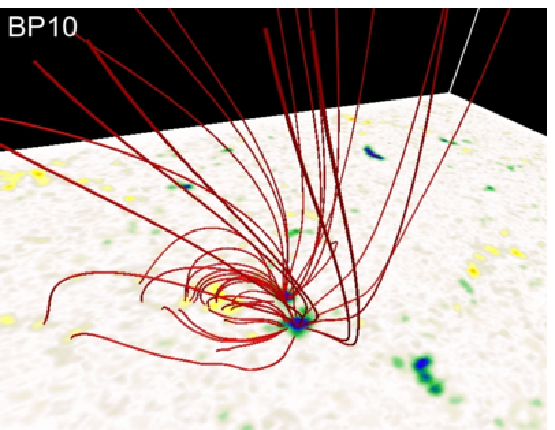}
\hspace{.1 cm}
\includegraphics[scale=.8]{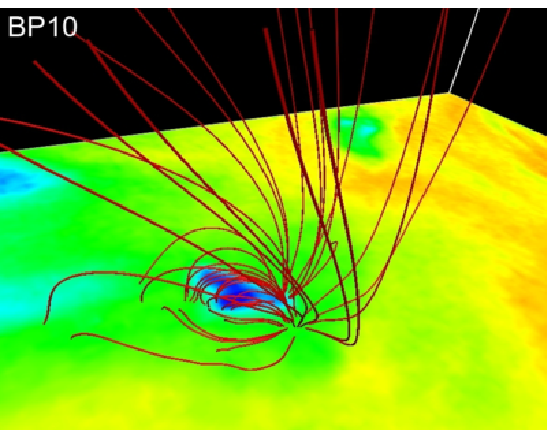}
}
\caption[]{Snapshots of QS BP8--BP10 as in \Fig{BPrest_CH.fig}. The times for the frames are BP8 -- 19:56:34~UT, BP9 -- 14:50:36~UT and BP10 -- 13:09:42~UT on their respective dates, \Tab{tab:bp_obs}. In BP8 and BP9 additional single coloured field lines are included to show the local field line structure near the projected X-ray regions Notice that for BP10 the base image in the right frame is the AIA 193. See the included animations for their time evolution.
\label{BPrest_QS.fig}
}}
\end{figure*}

\begin{figure*}[!t]
  \center
  \includegraphics[scale=.45]{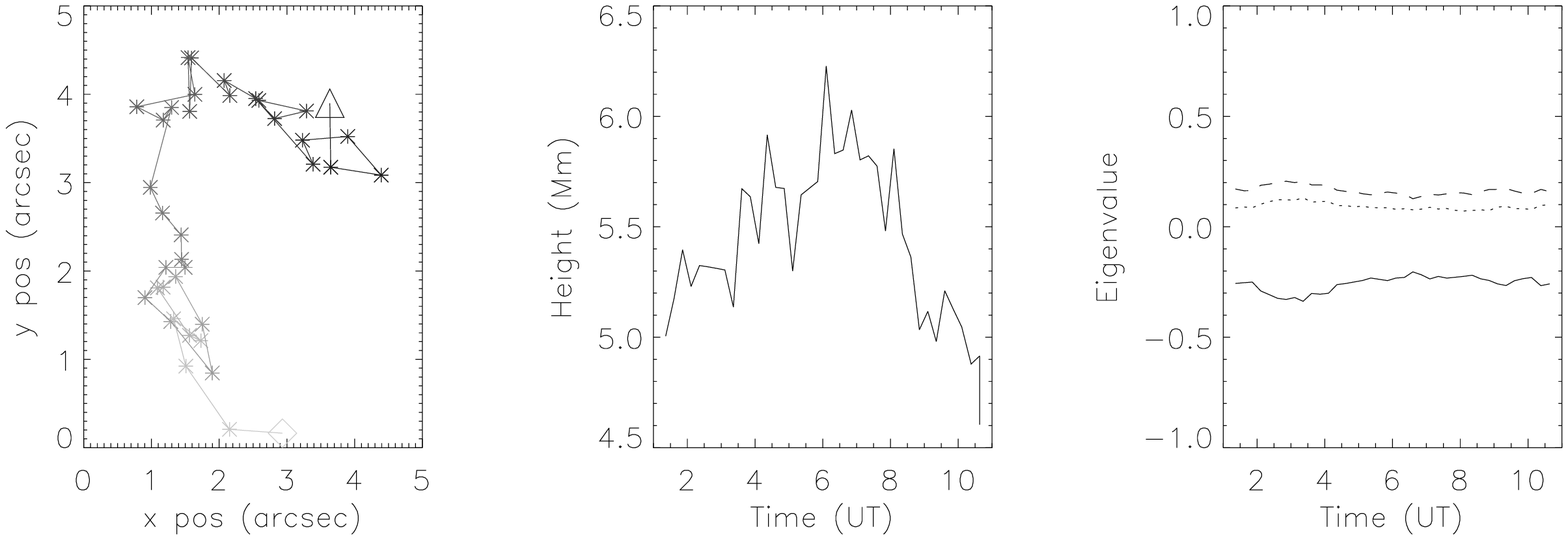}\\
  \includegraphics[scale=.45]{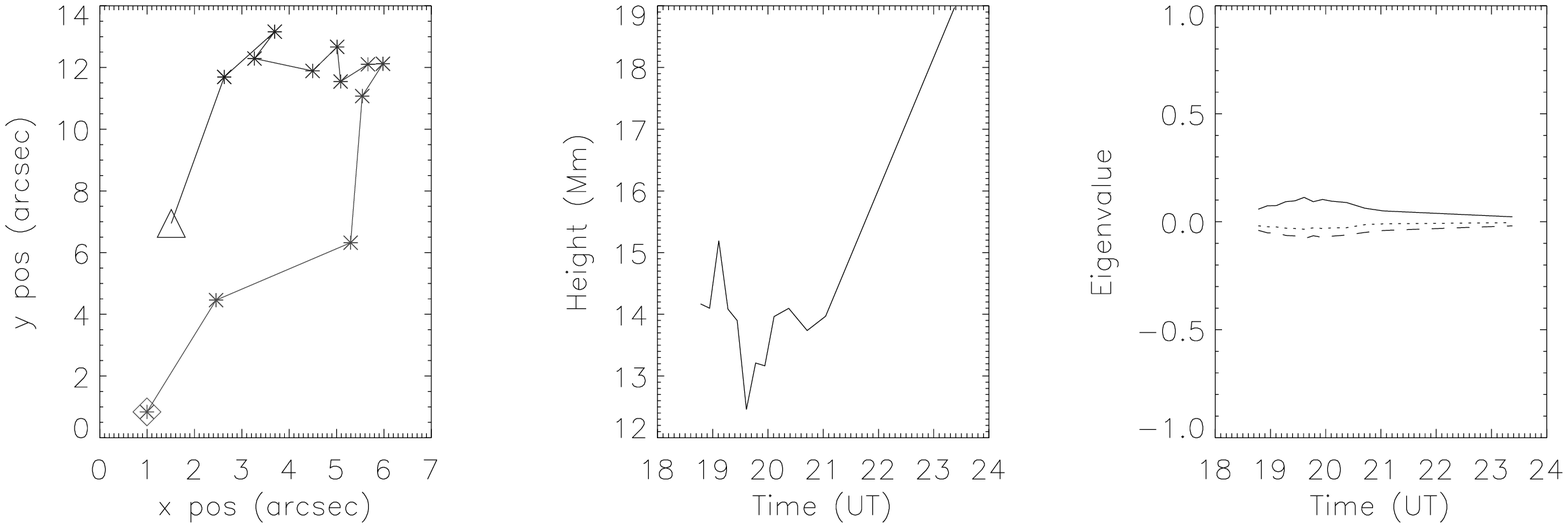} \\
  \includegraphics[scale=.45]{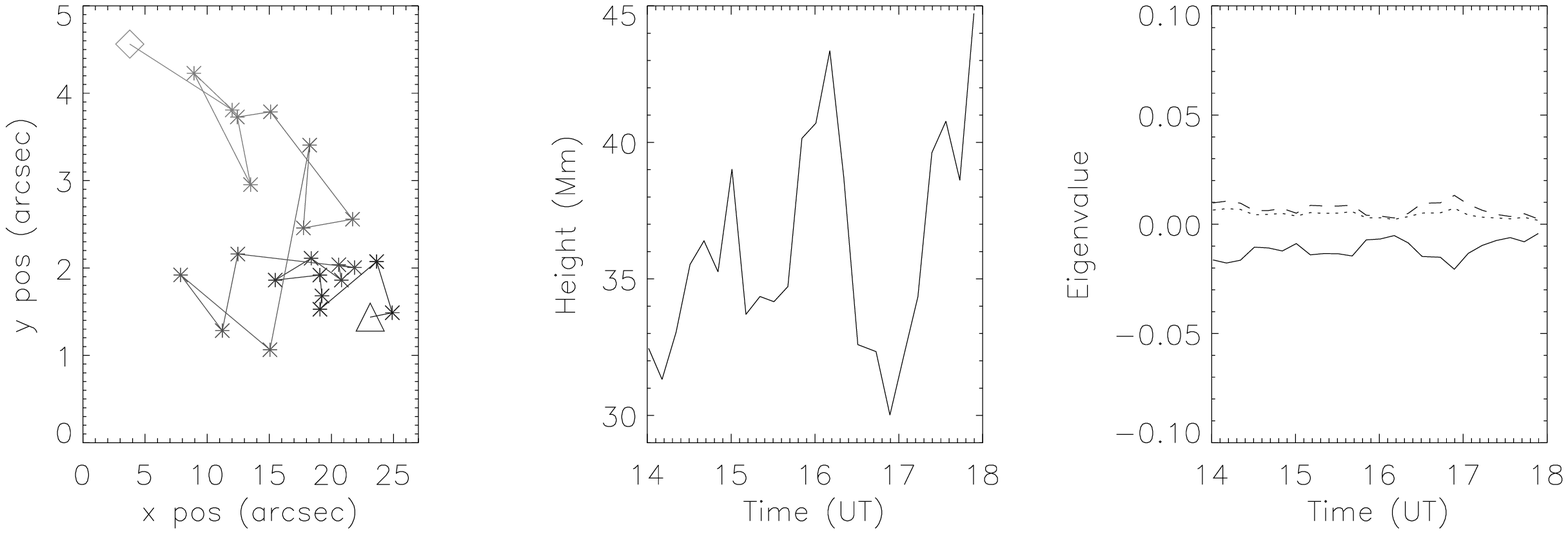} 
  \caption[]{The same as \Fig{BP_1_Nulls.fig} for BP7 (top row), BP8 (middle), and BP9 (bottom).}
\label{BP_3_Nulls.fig}
\end{figure*}

{\it BP8} (\Fig{BPrest_QS.fig}) is found in a QS region on 2007 October 10 and it is present throughout the whole observational period. This BP was extensively investigated by \citet{2011A&A...526A.134A} including detailed important information of the BP region that need not be repeated here. This BP has a double structure: it can be said to be composed of two BPs that are contained inside the fan dome boundaries of a null point located relatively high in the corona. They appear as two sets of loops that connect the minority polarity, with two dominating polarities on either side of the spine axis. Both BPs have a size of approximately 20\arcsec $\times$ 20\arcsec\ that slightly vary in time. One of the BPs is brighter in X-rays and its X-ray intensity increases with time as the two corresponding polarities drift towards each other. Towards the end of the time series the null point changes character, both being more extended in horizontal direction -- implying a near 2D magnetic field structure in this region -- while the null point region increases significantly in height, like being ejected upwards. There do not seem to be any clear changes in the magnetogram that can explain this behaviour. 
This evolution is only partly seen in \Fig{BP_3_Nulls.fig}, when the identification of the null point failed over this time period where the fast and significant changes took place. 

\citet{2011A&A...526A.134A} use the MPOLE code to obtain information about the 3D magnetic field structure and find that the main structure in the X-ray emission is well represented by their potential model, that is shown by the blue field line region seen in \Fig{BPrest_QS.fig}. The two investigations are in good agreement on the low lying field line structure of this region.  

\begin{figure}
{
\includegraphics[scale=.8]{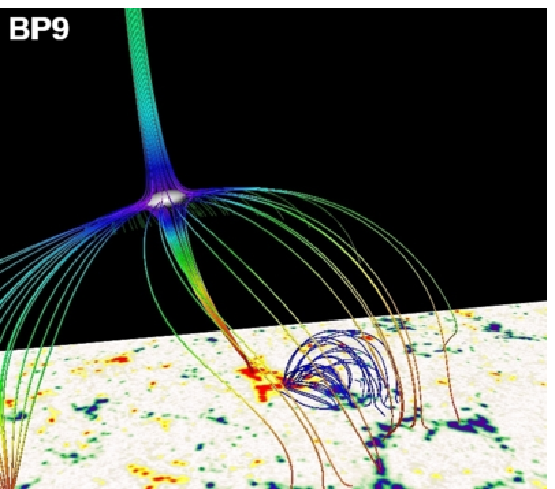}
\hfill
\includegraphics[scale=.8]{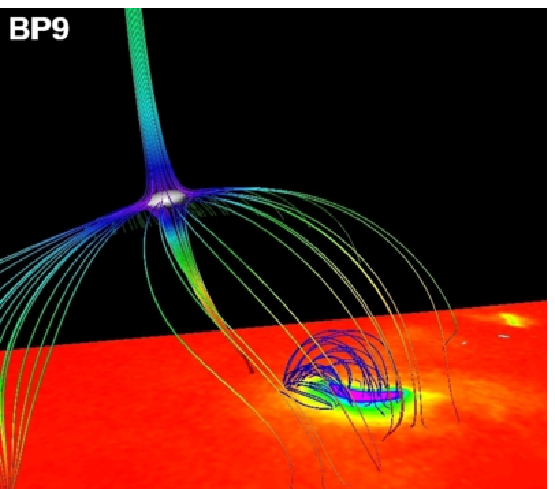}
}
\caption{Enlarged view of the BP9 case presented in \Fig{BPrest_QS.fig} showing the local field line structure around the concentrated X-ray emission. The left frame displays the surface magnetic field, while the right frame shows the associated X-ray emission at time 14:50:36~UT. See the included animation for the time evolution.
\label{BP9_local.fig}}
\end{figure}

{\it BP9} (\Fig{BPrest_CH.fig}) is found in a QS region on 2007 November 27 and is active for the full time of the observations. The BP has a single clear loop structure that shows minor variations over time. This BP is the largest of all studied cases with a size of $\approx$50\arcsec\ as measured from one end to the other of the loop structure. A null point is associated with this BP, but it is located very high above the BP region (\Fig{BP_3_Nulls.fig}). The null is located so high above the photosphere that the fan dome connects to the photosphere over a distance that is comparable to the smallest of the two dimensions of the observed SOT domain. 
The connectivity of the null is, therefore, defined by the periodicity of the extrapolation. The presence of the null point is due to a weak flux imbalance of the observed data, which may change if one cuts the magnetogram differently. The location and structure of the null point, therefore, cannot be determined with full confidence for this BP.
In this case the BP appears to be embedded in a bipolar field region located clearly inside the fan dome structure for the entire observing period (blue field lines in \Fig{BPrest_QS.fig}). This loop system at one end connects the structurally complicated minority flux concentration that anchors the spine axis and a number of dominant flux regions next to it (see \Fig{BP9_local.fig} for a zoom of the local field line structure). This causes the connectivity to have a significant divergence across this split in the dominant fluxes and thereby creates local quasi-separatrix layer \citep{1995JGR...10023443P} structures that may provide locations for local energy release when appropriately stressed. An inspection of the magnetogram time series shows the region between the strong dominant flux concentrations to contain many smaller flux concentrations that constantly move around adding to the general stressing of the local magnetic field.

{\it BP10} (\Fig{BPrest_CH.fig}) is found in a QS region on 2011 July 18 and is active for the full time of the SOT observations which unfortunately last only 57~min. The BP is clearly associated with a bipolar loop system with a size of $~$13\arcsec\ estimated from one visible footpoint to the other. There are no identified null points associated with the BP. This BP is, therefore, clearly different from the other identified BPs in the sample.

\section{Discussion}
\label{disc.sec}

Knowledge on the magnetic topology of coronal bright points throughout its formation, evolution and decay, is benchmarking for any further understanding on the physical mechanism that sustains this main building segment of the high temperature low corona. We explored the magnetic topology of 10 coronal BPs and found that 9 of these contain a magnetic null point located in the coronal domain. Seven of these are CH BPs, and two are QS BPs. The general structure of all null points is such that the spine axis, in one direction, connects to the minority polarity below the null point and, in the opposite direction, connects towards the top boundary of the domain. For the CH BPs the null point is located relative low in the atmosphere, and the X-ray emission is located either inside the fan-dome or in a fraction of the dome, mostly hosted in a dominant polarity flux concentration. Due to the low height of the null point, this strongly suggests that the null point is an important feature in the magnetic field for hosting the BP structure. For the QS BPs we see that the two identified null points sit at much higher in the atmosphere with the X-ray emission being located clearly inside the fan-dome region and associated with bipolar magnetic field regions showing a clear loop structure. Finally, BP10 does not show any relation to a coronal null point, and seems to be residing in a bipolar loop system.

 \citet{2010ApJ...714..130K} determined the heights of 210 BPs using a triangulation method on EUV images from the two STEREO spacecrafts in the 171~\AA, 195~\AA, 284~\AA\ and 304~\AA\ passbands. This provided them with heights for the BPs at different temperatures. For the 195~\AA\ filter they determine the BP mean height to be 6.7$\pm$2~Mm and determined lower heights for observations of BPs in passbands a with lower temperature response (171, 284 and 304~\AA). In comparison to this, \citet{2007AdSpR..39.1853T} used a linear field extrapolation of Kitt Peak magnetograms to investigate the alignment of the horizontal components of the extrapolated magnetic field at a given height with the emission contours of 10 BPs observed by EIT in the 195~\AA\ passband. They assumed a high correlation index to indicate the alignment of the emission with the loop summit and, therefore, can provide independent information about the height of the underlying BP loops. Their results show a scattering of BP heights within the 1--20 Mm range.
For all our CH cases, the null points are located relatively low in the atmosphere,  1--7 Mm, compared to the identified QS null points,  12--40 Mm.  The CH nulls are therefore inside the typical hight range of the observations, while the QS nulls are on the high side in comparison to the two investigations mentioned above. With the differences in our null point heights for the two solar regions,
the CH nulls will  have a much higher likelihood for being directly involved in the energy release process than for the QS BPs. This difference in height of the magnetic structures between the two regions is supported by the work of \citet{2004SoPh..225..227W}, where it is found that closed field lines are clearly lower in CH regions compared to QS regions. To investigate the importance of the null points in the different regions in relation to the BP regions, one has to conduct numerical 3D MHD modelling. To further check on the differences in magnetic structure between the two regions, more cases from QS regions must be studied in a comparable way. As mentioned above a dedicated observing campaign with SOT/EIS/Hinode and IRIS coupled with data from the AIA/SDO will be pursued.

In relation to this, \citet{2012ApJ...746...19Z} found that only two of the
13 bright points they selected in their region hosted a null point above the
BP, whereas the 11 remaining ones were located in general bipolar loop
systems. By inspection of full disk images from the day of their
observations, it is clear that these data correspond to a QS region and can
therefore be best compared with our BP8 to BP10 cases. Given the larger
number of QS BPs in their study we have tried to understand the possible
reasons for this discrepancy, in particular concerning the differences in the
underlying data and the treatment thereof. We do this in two parts:

\noindent a) a parameter that may influence the results is the spatial
resolution of the observational data that serve as a basis for the
extrapolation. \cite{2012ApJ...746...19Z} used longitudinal GONG magnetograms 
starting at midnight on 2007 March 16. 
From the header information of the corresponding data files (as
retrieved from the  GONG data base) we derived a pixel size of
those observations of 2.55\arcsec, to be compared with the pixel size of 
our SOT/NFI magnetograms (0.16\arcsec): SOT has $16$ times higher pixel resolution. 
Would the results presented in earlier sections of the present paper change, with null
point structures possibly vanishing, if we took as a basis lower-resolution
magnetograms?  To test for this, we have conducted an experiment where we
have degraded the SOT data to the GONG-magnetogram level by first convolving
with a Gaussian Point Spread Function (PSF) corresponding to the pixel size
ratio, followed by rebining to the pixel size of the GONG data. We have used
for that our 2007 Nov 9 initial data set, and carried out the degradation
process in four steps, i.e., degrading by pixel-size factors 2, 4, 8, and 16
to create intermediate-resolution cases.  For each of the new resolutions
new potential extrapolations were done, and a visual comparison of the different
resolutions were carried out.

\begin{figure*}[!ht]
\hfill \,
\includegraphics[scale=2.7]{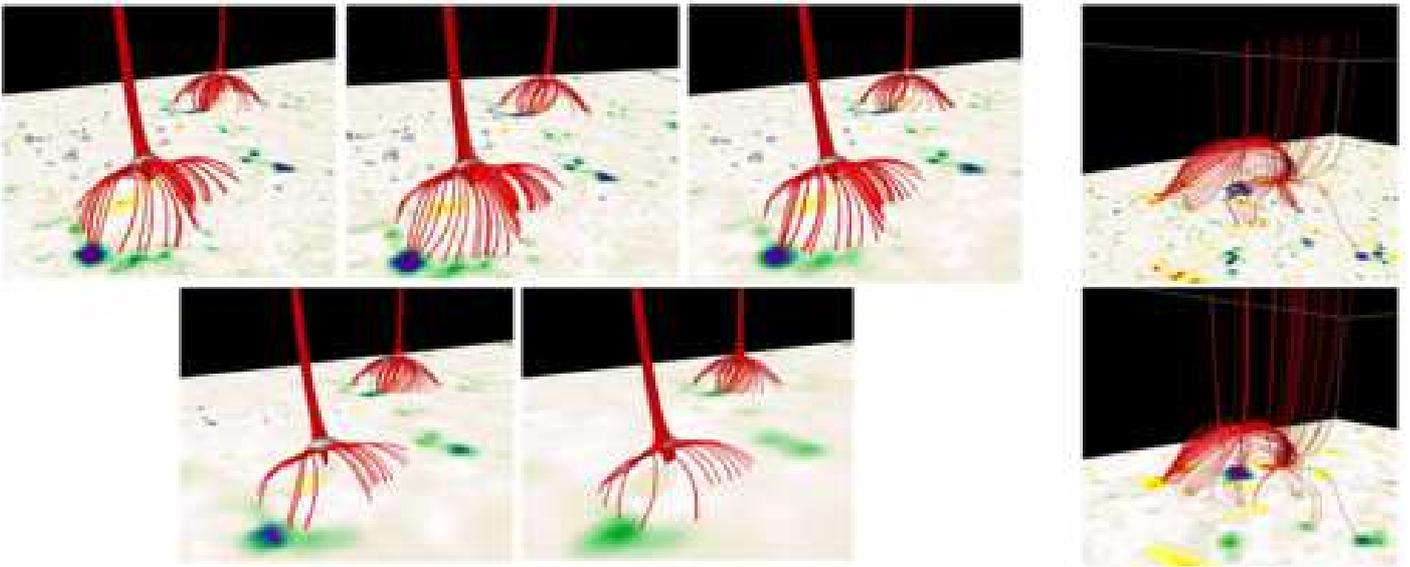}
\hfill \,

\caption[]{The frames show what happens when the pixel resolution of the
  magnetogram is degraded in a number of steps using a PSF convolution
  method. The left block of images show the extrapolations based on the
  original SOT data, and with a factor of 2, 4, 8 and 16 times degradation in
  spatial resolution for the BP1 and BP2 bright points. The right block of
  images shows the original SOT data and a factor of 16 times lower resolution
  for the BP8 bright point. The photospheric magnetograms are shown using a
  $\pm$200 Gauss range; following the standard biasing technique of the VAPOR
  package, the seeds for the field line tracing are planted at locations with
  very low values of the total field strength in a box stretching from the
  photosphere to several Mm height in the BP domain. 
\label{resolution_compare.fig}
}
\end{figure*}

\Fig{resolution_compare.fig} shows the five different resolutions with their
associated extrapolated field line configurations for two of the coronal hole
bright points, namely BP1 and BP2 (left block of frames), together with the
extreme (i.e., SOT original and fully degraded) resolutions for one of the
QS BPs, namely BP8 (right block of frames). In both cases we see that
the null point structure is maintained in spite of the severe degradation of
the magnetograms, as apparent in the surface distributions in the
figure. This is reassuring in terms of the robustness of the extrapolation
technique we are using to locate null points above photospheric regions with
mixed (in particular, embedded) polarities, especially those linked to
coronal bright points in as far as the latter have sizes larger than the pixel 
resolution of the magnetograms. 
On the other hand, the puzzle of the discrepancy between the global
conclusion of \citet{2012ApJ...746...19Z} and ours must be explored further,
for instance as follows: 

\noindent b) we have obtained GONG magnetogram data in the time range studied by
\citet{2012ApJ...746...19Z}, specifically, 2007 March 16, 00:01:16~UT, (as in
their Fig.~7) and carried out an extrapolation for a region of $900 \arcsec
\times 900 \arcsec$ that includes the domain with 13 bright points shown in
their Fig.~1. The magnetic structure in the neighbourhood of those BPs was
then re-investigated. Doing this we found that 8 BPs were clearly associated
with magnetic null points, 2 were structures of a different kind (akin to
bald patches), a further case was a simple bipolar loop configuration, and
there were two additional marginal cases where the identification with null
points was ambiguous. These results are significantly different to the
conclusions of \citet{2012ApJ...746...19Z}, who only found two null points:
we can only guess that the reason for the discrepancy lies with the specific
methods and techniques used to obtain the field lines, like, e.g., the region
chosen to carry out the extrapolation. We find that the average height above
the photosphere of the 8 clearly identified null points is $6$ Mm, with all
cases located at heights between $4$ and $10$ Mm.

To further explore this issue, additionally to the GONG data, we downloaded
the full-disk SOHO/MDI magnetogram at about the same time (01:39:30 ~UT) and
date, which has a  pixel size of $1.99$\arcsec. We
selected from it a domain basically coinciding with the one used above for
the GONG magnetogram and carried out a field extrapolation. The MDI
magnetogram has an excellent level of flux balance (around $2\%$ ratio of
total signed flux to total unsigned flux for our region, or about $3.5 \%$
for \citeauthor{2012ApJ...746...19Z}'s original region), much better than the
corresponding numbers for the GONG data. The peak field strength in the MDI
region, about $500$ G, is also much higher than that in the GONG magnetogram
(close to $200$G). Our region being roughly centered on disk center, we do
not expect the deviation between line-of-sight and vertical field components
to play any relevant role in reaching the conclusions below.  We could
clearly identify 8 null points associated with BPs in the list, with 6 of
them coinciding with positive identifications in our GONG extrapolation.
Two of the negative identifications were  found in both
extrapolations. Two cases were marginal cases in either extrapolation; for
the remaining three cases there is no clear explanation for the discrepancy,
except perhaps through the large differences in the two magnetograms
concerning field strength of the flux concentrations. The null points
identified in the MDI extrapolation are located at slightly higher positions
above the photosphere, namely between $6$ and $12$ Mm, with an average of $8$
Mm. 

All in all, we find that for a significant number of cases in the list of
\citet{2012ApJ...746...19Z} there is an association between a bright point and a null
point structure both in the GONG and the MDI datacheck. 

From the analysis of the CH BP regions, we can conclude that the presence of a magnetic null point above the BP region appears to be very important. In all cases the BP regions are associated with some fraction of the fan-dome that encloses the minority polarity flux. Magnetic null points are known to be a location that can host magnetic reconnection processes, and through this process change the connectivity of the magnetic field globally while affecting all of the dome region due to the flipping of magnetic field lines through the diffusion region. The basic evolution of an initially potential magnetic null point has been investigated in detail by \citet{2007PhPl...14e2106P}, \citet{2011A&A...529A..20G}, and \citet{2011A&A...534A...2G}, which shows how the null point  structure is perturbed strongly and generates systematic outflow regions that push the reconnected flux out of the current sheet region. A more realistic configuration is investigated by \citet{2013ApJ...774..154P}. It has a configuration that is comparable to the one shown in \Fig{extrapolation_compare.fig} where a flux balance of the data set is enforced. The  investigation in \citet{2013ApJ...774..154P}  shows how flux is transported through the null point while changing the connectivity patterns of the involved flux, and how different regions around the fan-dome change connectivity in the process. A similar evolution may take place in the CH BP magnetic field structures where the regions of change in connectivity is strongly altered by the inhomogeneity of the photospheric flux concentrations.

The skeleton defined by the null point located in the corona is very resilient and this structure may only be simplified by moving the null point out through the photospheric boundary. For the BPs discussed here the null points typically exist longer than the BPs are observed in X-rays. Only the BP4 case is different, where the X-ray emission is activated together with the formation of the null point through the photosphere and its following motion up into the coronal domain. As the X-ray emission vanishes, the null point vanished down through the photosphere again. This is the case that most clearly shows the importance of the null point in the appearance of the X-ray BP.
Having this characteristic magnetic skeleton raises the question about its role in the energy release process needed for heating the plasma to coronal temperatures. Looking at the time evolution of the magnetic field and the changes of the dome structures discussed in this paper (see the attached animations), it is clear that the convective motions constantly batter the footpoints of the magnetic field. 
Convection is a stochastic process, so how does this provide a long time heating of a localised region of the fan-dome? The motions in the previous mentioned numerical investigations have all been imposing a systematic stressing of the null point system. So we may not directly infer these results to the BP systems. From the extrapolated field models shown in \Fig{BPrest_CH.fig} and \Fig{BPrest_QS.fig} one sees the X-ray emission, in many cases, to be associated with a dominating flux concentration, while there also exist examples where the emission is located above a weak field region (BP7). How do we explain how these different magnetic structured regions can host X-ray emission? 
The dynamical response of the complicated magnetic field region to convective motions cannot be predicted from the knowledge of the magnetic field structure alone. To get a qualified understanding of the effect of the imposed convective motions, we have to turn to dedicated numerical experiments. Such work is presently in progress \citep{Moreno-Insertis-Galsgaard-2017}, where investigations of large scale systematic stressing is being studied, but we also need to explore the stochastic stressing process to establish the effect of these on the magnetic structure. 
  
 The plasma cooling time due to radiation (and conduction) of the BP X-ray plasma is on the order of 5~min \citep[See their Table~2]{2011A&A...526A.134A}, which indicates that a continuous energy input to the region is required. This energy may arise in part from free magnetic energy in the magnetic structure and from a continues stressing of the system. One may argue, therefore, that a potential field extrapolation is not the best choice when describing the magnetic field structure of BP regions. Presently though no sufficiently high quality vector magnetograms exist outside active regions that can be used for more advanced extrapolation methods like non-linear force free field extrapolations.  
In addition, the requirement of the more advance methods is magnetic flux balance over the area of the adopted magnetogram. A condition that is never fulfilled in CH regions. Please see also the discussion in Section~2.2 on the comparison of magneto-static and potential field extrapolation methods.

\citet{2008A&A...492..575P} and \citet{2011A&A...526A.134A} investigated a few BP regions using the MPOLE code.
This code uses a relative low number of point sources to represent a magnetogram and from this analytic description they identify magnetic null points in the photosphere and their connections through spine axis and fan planes. 
This reduced information was used to investigate the structure of the 3D magnetic field and to relate the magnetic structures to the BPs'  X-ray  emission. Both studies find the BP magnetic field structures to be bipolar in nature from identifying field lines that align with the X-ray emission. During this process they do not seem to investigate the large scale structure of the magnetic field, which prevents them from identifying any potential null points above the BP region. A direct comparison between their and our results on these scales is therefore not possible. 

\section{Summary and Conclusions}
\label{conc.sec}
In this paper we have used SOT/XRT/Hinode and HMI/AIA/SDO observations to investigate the topology of X-ray/EUV BPs. 3D magnetic fields have been derived for different times using a potential field extrapolation for the full size of the observed magnetograms. Within these data the location of the BP regions were identified and the structure of the associated magnetic field investigated. The topology of the local magnetic field was visualised using the information of the position of the null points associated with the identified BP regions. This was used to investigate the close relation between the magnetic field structure and the projected location of the X-ray emission for the 10 selected BP regions.  

The topology investigation showed a close relation between the CH BPs X-ray emission and the structure of the magnetic field defined by 3D null points. This shows that for all CH cases analysed here, the null points fan-dome structure embraces the X-ray emission in slightly different ways for the observed BPs. This strongly suggests that the presence of the null-point region located in the coronal region is crucial for the energy release processes for these BP regions.  

Two cases of QS BPs also shows the presence of 3D null points, but these are located much higher in the atmosphere compared to the CH null points, while the third case has no coronal null point. This height of the two null points above the photosphere and the lack of knowledge of the height of the X-ray emission region makes it less clear how important the null points are for these BP regions. Just making field line traces from the photosphere makes one to assume that it is the low lying field lines, that show a bipolar structure, that represents the important structure of the BP regions. A good indication of the height of the X-ray emitting region would be important for identifying which part of the 3D magnetic field that hosts it. This would then allow for a discrimination between the different scenarios for the local magnetic field line structures responsible for hosting the BP regions and provide a substantial clue towards the heating process.  

Ignoring the problem with the height of the X-ray region, and simply assuming that the simple approach for the local field line structure provides a sufficient representation for the structure of the field, then the investigated data indicate that two possible mechanisms exist for powering the BP regions. One which mainly works for the CH BPs where the coronal null point is the key location for the energy conversion. A different mechanism may represent the QS BPs where the local bipolar loop structure is more important than the presence of the associated null-point region. More observational data with high spacial resolution for QS BP are needed to better understand what the correct magnetic structures for this type of BP regions are. When this is determined more conclusively a much better starting point for investigating the dynamical evolution of the BPs will be presented.

\begin{acknowledgements}
KG acknowledges support through a research grant (VKR023406) from
VILLUMFONDEN.  FMI is grateful for the financial
support by the Spanish Ministry of Economy and Competitiveness (MINECO)
through project AYA2014-55078-P. Z.H. acknowledges NSFC for financial support
via grant 41404135. TW acknowledges DLR-grant 50 OC 1301 and
DFG-grant WI 3211/4-1. The authors are grateful for the use of the 
softwareImagery produced by VAPOR (www.vapor.ucar.edu), 
a product of the Computational Information Systems Laboratory at the National
Center for Atmospheric Research.  
\end{acknowledgements}

\bibliography{ref}

\begin{thebibliography}{46}
\expandafter\ifx\csname natexlab\endcsname\relax\def\natexlab#1{#1}\fi

\bibitem[{{Alexander} {et~al.}(2011){Alexander}, {Del Zanna}, \&
  {Maclean}}]{2011A&A...526A.134A}
{Alexander}, C.~E., {Del Zanna}, G., \& {Maclean}, R.~C. 2011, \aap, 526, A134

\bibitem[{{Brown} {et~al.}(2001){Brown}, {Parnell}, {Deluca}, {Golub}, \&
  {McMullen}}]{2001SoPh..201..305B}
{Brown}, D.~S., {Parnell}, C.~E., {Deluca}, E.~E., {Golub}, L., \& {McMullen},
  R.~A. 2001, \solphys, 201, 305

\bibitem[{{Galsgaard} \& {Pontin}(2011{\natexlab{a}})}]{2011A&A...534A...2G}
{Galsgaard}, K. \& {Pontin}, D.~I. 2011{\natexlab{a}}, \aap, 534, A2

\bibitem[{{Galsgaard} \& {Pontin}(2011{\natexlab{b}})}]{2011A&A...529A..20G}
{Galsgaard}, K. \& {Pontin}, D.~I. 2011{\natexlab{b}}, \aap, 529, A20

\bibitem[{{Golub} {et~al.}(2007){Golub}, {Deluca}, {Austin}, {Bookbinder},
  {Caldwell}, {Cheimets}, {Cirtain}, {Cosmo}, {Reid}, {Sette}, {Weber},
  {Sakao}, {Kano}, {Shibasaki}, {Hara}, {Tsuneta}, {Kumagai}, {Tamura},
  {Shimojo}, {McCracken}, {Carpenter}, {Haight}, {Siler}, {Wright}, {Tucker},
  {Rutledge}, {Barbera}, {Peres}, \& {Varisco}}]{2007SoPh..243...63G}
{Golub}, L., {Deluca}, E., {Austin}, G., {et~al.} 2007, \solphys, 243, 63

\bibitem[{{Golub} {et~al.}(1977){Golub}, {Krieger}, {Harvey}, \&
  {Vaiana}}]{1977SoPh...53..111G}
{Golub}, L., {Krieger}, A.~S., {Harvey}, J.~W., \& {Vaiana}, G.~S. 1977,
  \solphys, 53, 111

\bibitem[{{Golub} {et~al.}(1974){Golub}, {Krieger}, {Silk}, {Timothy}, \&
  {Vaiana}}]{1974ApJ...189L..93G}
{Golub}, L., {Krieger}, A.~S., {Silk}, J.~K., {Timothy}, A.~F., \& {Vaiana},
  G.~S. 1974, \apjl, 189, L93

\bibitem[{{Golub} {et~al.}(1976{\natexlab{a}}){Golub}, {Krieger}, \&
  {Vaiana}}]{1976SoPh...49...79G}
{Golub}, L., {Krieger}, A.~S., \& {Vaiana}, G.~S. 1976{\natexlab{a}}, \solphys,
  49, 79

\bibitem[{{Golub} {et~al.}(1976{\natexlab{b}}){Golub}, {Krieger}, \&
  {Vaiana}}]{1976SoPh...50..311G}
{Golub}, L., {Krieger}, A.~S., \& {Vaiana}, G.~S. 1976{\natexlab{b}}, \solphys,
  50, 311

\bibitem[{{Habbal}(1992)}]{1992AnGeo..10...34H}
{Habbal}, S.~R. 1992, Annales Geophysicae, 10, 34

\bibitem[{{Habbal} {et~al.}(1990){Habbal}, {Withbroe}, \&
  {Dowdy}}]{1990ApJ...352..333H}
{Habbal}, S.~R., {Withbroe}, G.~L., \& {Dowdy}, Jr., J.~F. 1990, \apj, 352, 333

\bibitem[{{Harvey} {et~al.}(1999){Harvey}, {Jones}, {Schrijver}, \&
  {Penn}}]{1999SoPh..190...35H}
{Harvey}, K.~L., {Jones}, H.~P., {Schrijver}, C.~J., \& {Penn}, M.~J. 1999,
  \solphys, 190, 35

\bibitem[{{Haynes} \& {Parnell}(2007)}]{2007PhPl...14h2107H}
{Haynes}, A.~L. \& {Parnell}, C.~E. 2007, Physics of Plasmas, 14, 082107

\bibitem[{{Huang} {et~al.}(2012){Huang}, {Madjarska}, {Doyle}, \&
  {Lamb}}]{2012A&A...548A..62H}
{Huang}, Z., {Madjarska}, M.~S., {Doyle}, J.~G., \& {Lamb}, D.~A. 2012, \aap,
  548, A62

\bibitem[{{Ichimoto} {et~al.}(2008){Ichimoto}, {Lites}, {Elmore}, {Suematsu},
  {Tsuneta}, {Katsukawa}, {Shimizu}, {Shine}, {Tarbell}, {Title}, {Kiyohara},
  {Shinoda}, {Card}, {Lecinski}, {Streander}, {Nakagiri}, {Miyashita},
  {Noguchi}, {Hoffmann}, \& {Cruz}}]{2008SoPh..249..233I}
{Ichimoto}, K., {Lites}, B., {Elmore}, D., {et~al.} 2008, \solphys, 249, 233

\bibitem[{{Kwon} {et~al.}(2012){Kwon}, {Chae}, {Davila}, {Zhang}, {Moon},
  {Poomvises}, \& {Jones}}]{2012ApJ...757..167K}
{Kwon}, R.-Y., {Chae}, J., {Davila}, J.~M., {et~al.} 2012, \apj, 757, 167

\bibitem[{{Kwon} {et~al.}(2010){Kwon}, {Chae}, \&
  {Zhang}}]{2010ApJ...714..130K}
{Kwon}, R.-Y., {Chae}, J., \& {Zhang}, J. 2010, \apj, 714, 130

\bibitem[{{Lemen} {et~al.}(2012){Lemen}, {Title}, {Akin}, {Boerner}, {Chou},
  {Drake}, {Duncan}, {Edwards}, {Friedlaender}, {Heyman}, {Hurlburt}, {Katz},
  {Kushner}, {Levay}, {Lindgren}, {Mathur}, {McFeaters}, {Mitchell}, {Rehse},
  {Schrijver}, {Springer}, {Stern}, {Tarbell}, {Wuelser}, {Wolfson}, {Yanari},
  {Bookbinder}, {Cheimets}, {Caldwell}, {Deluca}, {Gates}, {Golub}, {Park},
  {Podgorski}, {Bush}, {Scherrer}, {Gummin}, {Smith}, {Auker}, {Jerram},
  {Pool}, {Soufli}, {Windt}, {Beardsley}, {Clapp}, {Lang}, \&
  {Waltham}}]{2012SoPh..275...17L}
{Lemen}, J.~R., {Title}, A.~M., {Akin}, D.~J., {et~al.} 2012, \solphys, 275, 17

\bibitem[{{Longcope}(1996)}]{1996SoPh..169...91L}
{Longcope}, D.~W. 1996, \solphys, 169, 91

\bibitem[{{Longcope} \& {Parnell}(2009)}]{2009SoPh..254...51L}
{Longcope}, D.~W. \& {Parnell}, C.~E. 2009, \solphys, 254, 51

\bibitem[{{Low}(1991)}]{low91}
{Low}, B.~C. 1991, \apj, 370, 427

\bibitem[{{Madjarska} {et~al.}(2003){Madjarska}, {Doyle}, {Teriaca}, \&
  {Banerjee}}]{2003A&A...398..775M}
{Madjarska}, M.~S., {Doyle}, J.~G., {Teriaca}, L., \& {Banerjee}, D. 2003,
  \aap, 398, 775

\bibitem[{{Moreno-Insertis} \&
  {Galsgaard}(2017)}]{Moreno-Insertis-Galsgaard-2017}
{Moreno-Insertis}, F. \& {Galsgaard}. 2017, work in progress

\bibitem[{{Mou} {et~al.}(2016){Mou}, {Huang}, {Xia}, {Madjarska}, {Li}, {Fu},
  {Jiao}, \& {Hou}}]{2016ApJ...818....9M}
{Mou}, C., {Huang}, Z., {Xia}, L., {et~al.} 2016, \apj, 818, 9

\bibitem[{{Parnell} {et~al.}(1997){Parnell}, {Neukirch}, {Smith}, \&
  {Priest}}]{1997GApFD..84..245P}
{Parnell}, C.~E., {Neukirch}, T., {Smith}, J.~M., \& {Priest}, E.~R. 1997,
  Geophysical and Astrophysical Fluid Dynamics, 84, 245

\bibitem[{{Parnell} {et~al.}(1994{\natexlab{a}}){Parnell}, {Priest}, \&
  {Golub}}]{1994SoPh..151...57P}
{Parnell}, C.~E., {Priest}, E.~R., \& {Golub}, L. 1994{\natexlab{a}}, \solphys,
  151, 57

\bibitem[{{Parnell} {et~al.}(1994{\natexlab{b}}){Parnell}, {Priest}, \&
  {Titov}}]{1994SoPh..153..217P}
{Parnell}, C.~E., {Priest}, E.~R., \& {Titov}, V.~S. 1994{\natexlab{b}},
  \solphys, 153, 217

\bibitem[{{P{\'e}rez-Su{\'a}rez} {et~al.}(2008){P{\'e}rez-Su{\'a}rez},
  {Maclean}, {Doyle}, \& {Madjarska}}]{2008A&A...492..575P}
{P{\'e}rez-Su{\'a}rez}, D., {Maclean}, R.~C., {Doyle}, J.~G., \& {Madjarska},
  M.~S. 2008, \aap, 492, 575

\bibitem[{{Pontin} {et~al.}(2007){Pontin}, {Bhattacharjee}, \&
  {Galsgaard}}]{2007PhPl...14e2106P}
{Pontin}, D.~I., {Bhattacharjee}, A., \& {Galsgaard}, K. 2007, Physics of
  Plasmas, 14, 052106

\bibitem[{{Pontin} {et~al.}(2013){Pontin}, {Priest}, \&
  {Galsgaard}}]{2013ApJ...774..154P}
{Pontin}, D.~I., {Priest}, E.~R., \& {Galsgaard}, K. 2013, \apj, 774, 154

\bibitem[{{Pre{\'s} } \& {Phillips}(1999)}]{1999ApJ...510L..73P}
{Pre{\'s} }, P. \& {Phillips}, K.~H.~J. 1999, \apjl, 510, L73

\bibitem[{{Priest} \& {D{\'e}moulin}(1995)}]{1995JGR...10023443P}
{Priest}, E.~R. \& {D{\'e}moulin}, P. 1995, \jgr, 100, 23443

\bibitem[{{Priest} {et~al.}(1994){Priest}, {Parnell}, \&
  {Martin}}]{1994ApJ...427..459P}
{Priest}, E.~R., {Parnell}, C.~E., \& {Martin}, S.~F. 1994, \apj, 427, 459

\bibitem[{{Tian} {et~al.}(2007){Tian}, {Tu}, {He}, \&
  {Marsch}}]{2007AdSpR..39.1853T}
{Tian}, H., {Tu}, C.-Y., {He}, J.-S., \& {Marsch}, E. 2007, Advances in Space
  Research, 39, 1853

\bibitem[{{Tsuneta} {et~al.}(2008){Tsuneta}, {Ichimoto}, {Katsukawa}, {Nagata},
  {Otsubo}, {Shimizu}, {Suematsu}, {Nakagiri}, {Noguchi}, {Tarbell}, {Title},
  {Shine}, {Rosenberg}, {Hoffmann}, {Jurcevich}, {Kushner}, {Levay}, {Lites},
  {Elmore}, {Matsushita}, {Kawaguchi}, {Saito}, {Mikami}, {Hill}, \&
  {Owens}}]{2008SoPh..249..167T}
{Tsuneta}, S., {Ichimoto}, K., {Katsukawa}, Y., {et~al.} 2008, \solphys, 249,
  167

\bibitem[{{Ugarte-Urra} {et~al.}(2004){Ugarte-Urra}, {Doyle}, {Madjarska}, \&
  {O'Shea}}]{2004A&A...418..313U}
{Ugarte-Urra}, I., {Doyle}, J.~G., {Madjarska}, M.~S., \& {O'Shea}, E. 2004,
  \aap, 418, 313

\bibitem[{{Vaiana} {et~al.}(1973){Vaiana}, {Krieger}, \&
  {Timothy}}]{1973SoPh...32...81V}
{Vaiana}, G.~S., {Krieger}, A.~S., \& {Timothy}, A.~F. 1973, \solphys, 32, 81

\bibitem[{{van Ballegooijen} \& {Martens}(1989)}]{1989ApJ...343..971V}
{van Ballegooijen}, A.~A. \& {Martens}, P.~C.~H. 1989, \apj, 343, 971

\bibitem[{{von Rekowski} {et~al.}(2006{\natexlab{a}}){von Rekowski}, {Parnell},
  \& {Priest}}]{2006MNRAS.366..125V}
{von Rekowski}, B., {Parnell}, C.~E., \& {Priest}, E.~R. 2006{\natexlab{a}},
  \mnras, 366, 125

\bibitem[{{von Rekowski} {et~al.}(2006{\natexlab{b}}){von Rekowski}, {Parnell},
  \& {Priest}}]{2006MNRAS.369...43V}
{von Rekowski}, B., {Parnell}, C.~E., \& {Priest}, E.~R. 2006{\natexlab{b}},
  \mnras, 369, 43

\bibitem[{{Webb} {et~al.}(1993){Webb}, {Martin}, {Moses}, \&
  {Harvey}}]{1993SoPh..144...15W}
{Webb}, D.~F., {Martin}, S.~F., {Moses}, D., \& {Harvey}, J.~W. 1993, \solphys,
  144, 15

\bibitem[{{Wiegelmann} {et~al.}(2015){Wiegelmann}, {Neukirch}, {Nickeler},
  {Solanki}, {Mart{\'{\i}}nez Pillet}, \& {Borrero}}]{2015ApJ...815...10W}
{Wiegelmann}, T., {Neukirch}, T., {Nickeler}, D.~H., {et~al.} 2015, \apj, 815,
  10

\bibitem[{{Wiegelmann} \& {Solanki}(2004)}]{2004SoPh..225..227W}
{Wiegelmann}, T. \& {Solanki}, S.~K. 2004, \solphys, 225, 227

\bibitem[{{Zhang} {et~al.}(2001){Zhang}, {Kundu}, \&
  {White}}]{2001SoPh..198..347Z}
{Zhang}, J., {Kundu}, M.~R., \& {White}, S.~M. 2001, \solphys, 198, 347

\bibitem[{{Zhang} {et~al.}(2012){Zhang}, {Chen}, {Guo}, {Fang}, \&
  {Ding}}]{2012ApJ...746...19Z}
{Zhang}, Q.~M., {Chen}, P.~F., {Guo}, Y., {Fang}, C., \& {Ding}, M.~D. 2012,
  \apj, 746, 19

\bibitem[{{Zwaan}(1987)}]{1987ARA&A..25...83Z}
{Zwaan}, C. 1987, \araa, 25, 83

\end{thebibliography}

\begin{appendix}
\section{Online material}

Animations for the 10 BPs discussed in the paper above. 

\begin{figure}[ht]
{ \hfill
\includegraphics[scale=1.1]{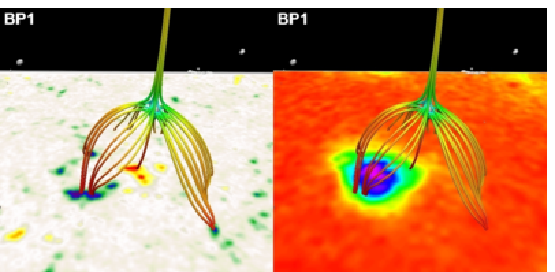}
\hfill}
\caption{BP1, showing a combination of field line traces and the underlying magnetogram and the X-ray emission as a function of time. }
\label{movie1.fig}
\end{figure}

\begin{figure}[ht]
{\hfill
\includegraphics[scale=1.1]{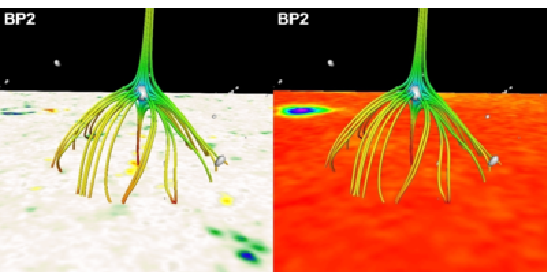}
\hfill}
\caption{BP2, showing a combination of field line traces and the underlying magnetogram and the X-ray emission as a function of time. }
\label{movie2.fig}
\end{figure}

\begin{figure}[ht]
{\hfill
\includegraphics[scale=1.1]{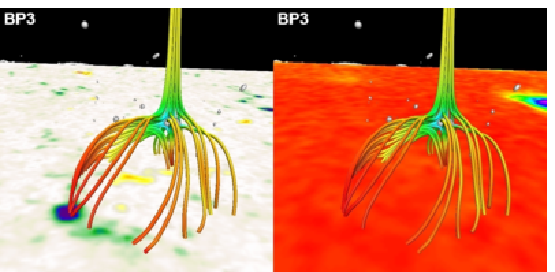}
\hfill}
\caption{BP3, showing a combination of field line traces and the underlying magnetogram and the X-ray emission as a function of time. }
\label{movie3.fig}
\end{figure}

\begin{figure}[ht]
{\hfill
\includegraphics[scale=1.1]{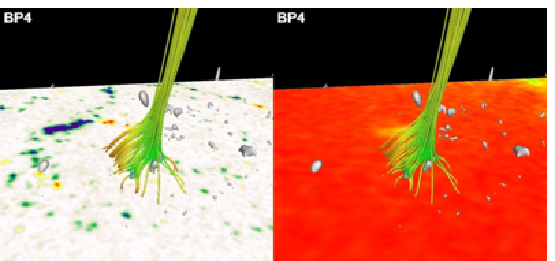}
\hfill}
\caption{BP4, showing a combination of field line traces and the underlying magnetogram and the X-ray emission as a function of time. }
\label{movie4.fig}
\end{figure}

\begin{figure}[ht]
{\hfill
\includegraphics[scale=1.1]{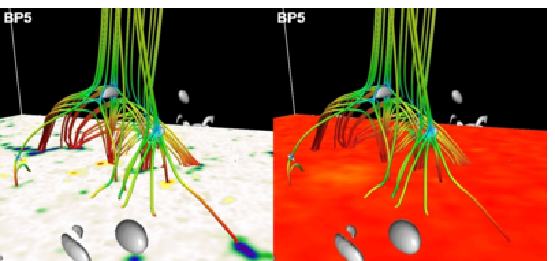}
\hfill}
\caption{BP5, showing a combination of field line traces and the underlying magnetogram and the X-ray emission as a function of time. }
\label{movie5.fig}
\end{figure}

\begin{figure}[ht]
{\hfill
\includegraphics[scale=1.1]{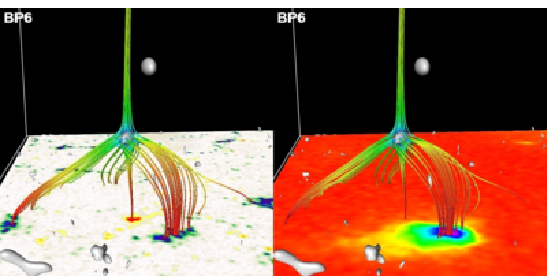}
\hfill}
\caption{BP6, showing a combination of field line traces and the underlying magnetogram and the X-ray emission as a function of time. }
\label{movie6.fig}
\end{figure}

\begin{figure}[ht]
{\hfill
\includegraphics[scale=1.1]{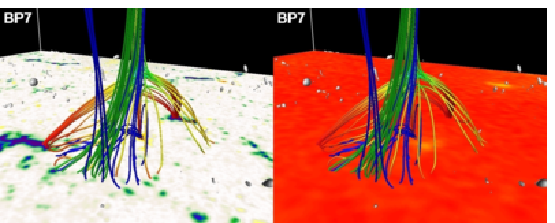}
\hfill}
\caption{BP1, showing a combination of field line traces and the underlying magnetogram and the X-ray emission as a function of time. }
\label{movie7.fig}
\end{figure}

\begin{figure}[ht]
{\hfill
\includegraphics[scale=1.1]{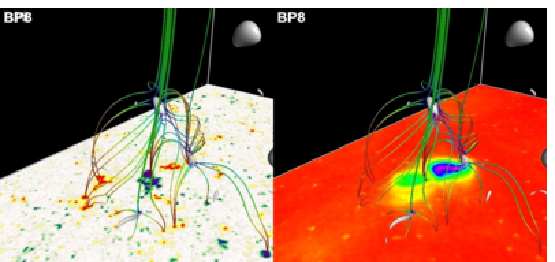}
\hfill}
\caption{BP8, showing a combination of field line traces and the underlying magnetogram and the X-ray emission as a function of time. }
\label{movie8.fig}
\end{figure}

\begin{figure}[ht]
{\hfill
\includegraphics[scale=1.1]{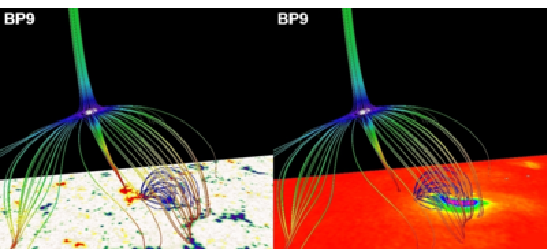}
\hfill}
\caption{BP9, showing a combination of field line traces and the underlying magnetogram and the X-ray emission as a function of time. }
\label{movie9.fig}
\end{figure}

\begin{figure}[ht]
{\hfill
\includegraphics[scale=1.1]{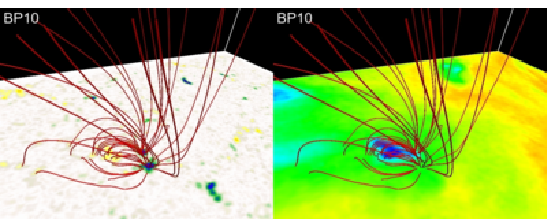}
\hfill}
\caption{BP10, showing a combination of field line traces and the underlying magnetogram and the AIA193 emission as a function of time. }
\label{movie10.fig}
\end{figure}
\end{appendix}

\end{document}